\documentclass[10pt]{bmc_article}

\usepackage{cite} 
\usepackage{url}  
\usepackage{ifthen}  
\usepackage{multicol}   
\usepackage[utf8]{inputenc} 
\usepackage{graphicx}
\usepackage{cite}
\usepackage[table]{xcolor}
\usepackage{booktabs}
\usepackage{multirow}
\usepackage{amsmath,amssymb}
\usepackage{pstricks}
\usepackage{amsmath,amssymb,subfigure,epsfig,bbm,stmaryrd,MnSymbol,mathrsfs}
\usepackage{subfigure}

\DeclareMathAlphabet{\mathpzc}{OT1}{pzc}{m}{it}

\newboolean{publ}

\newenvironment{bmcformat}{\fussy\setboolean{publ}{true}}{\fussy}

\begin{document}
\begin{bmcformat}

\title{A Hybrid Global Minimization Scheme for Accurate Source Localization in Sensor Networks}

\author{Hamidreza Aghasi$^1$%
         \email{Hamidreza Aghasi - aghasi@ee.sharif.edu}\!\!
       \and
         Hamidreza Amindavar$^2$%
         \email{Hamidreza Amindavar - hamidami@aut.ac.ir}%
         \;and
         Alireza Aghasi$^3$%
         \email{Alireza Aghasi - aaghas01@ece.tufts.edu}%
      }


\address{%
    \iid(1)Department of Electrical Engineering, Sharif University of Technology, Tehran, Iran\\
    \iid(2)Department of Electrical Engineering, Amirkabir University of Technology, Tehran, Iran\\
    \iid(3)Department of Electrical and Computer Engineering, Tufts
    University, Medford, MA, USA
}%

\maketitle


\begin{abstract}
We consider the localization problem of multiple wideband sources
in a multi-path environment by coherently taking into account the
attenuation characteristics and the time delays in the reception
of the signal. Our proposed method leaves the space for
unavailability of an accurate signal attenuation model in the
environment by considering the model as an unknown function with
reasonable prior assumptions about its functional space. Such
approach is capable of enhancing the localization performance
compared to only utilizing the signal attenuation information or
the time delays. In this paper, the localization problem is
modeled as a cost function in terms of the source locations,
attenuation model parameters and the multi-path parameters. To
globally perform the minimization, we propose a hybrid algorithm
combining the differential evolution algorithm with the
Levenberg-Marquardt algorithm. Besides the proposed combination of
optimization schemes, supporting the technical details such as
closed forms of cost function sensitivity matrices are  provided.
Finally, the validity of the proposed method is examined in
several localization scenarios, taking into account the noise in
the environment, the multi-path phenomenon and considering the
sensors not being synchronized.
\end{abstract}

\ifthenelse{\boolean{publ}}{\begin{multicols}{2}}{}
\section{Introduction}
A challenging and highly demanding signal processing application
is the localization of signal sources using the physical
measurements at some sensors in the environment. Source
localization has become an important task in various applications
such as mobile communications, global positioning system (GPS),
radar, sonar, navigation, seismology and geophysics
\cite{akyildiz2002wireless,niculescu2001ad, elfes1987sonar,
savarese2001location, hart2006environmental}.

During the recent decades various algorithms have been proposed to
estimate the location of the signal sources. These methods utilize
different signal characteristics at different sensors and
generally can be classified in three main categories: using the
time difference of arrival (TDOA); analyzing the signal direction
of arrival (DOA) at distinct arrays; and using the differences in
the signal amplitude or received energy level. For a constant
propagation speed, the TDOA among different sensors is
proportional to the source-sensor range differences and may be
estimated through methods such as cross correlation (CC)
\cite{carter1993coherence} or its generalized version (GCC)
\cite{knapp1976generalized}. The source locations can then be
estimated using geometric methods such as linear, spherical or
hyperbolic intersections \cite{brandstein1997closed,
smith1987closed, chan1994simple}. To estimate the DOA, for
narrowband signals, high resolution algorithms such as multiple
signal classification (MUSIC) \cite{schmidt1986multiple} and
maximum likelihood (ML) \cite{ziskind1988maximum} are proposed. In
\cite{chen2002maximum}, the authors propose an approximate maximum
likelihood method (AML) for wideband signals using spectral
properties of the signal when rather long sample streams are
available. In this method, the corresponding cost function can be
directly expressed in terms of the source locations or in a far
field case may be expressed in terms of the relative time delays
followed by a post processing step to find the source locations
from the corresponding DOAs. The post processing step may be
carried out through geometric methods such as cross bearing or a
machine learning approach such as the support vector machine (SVM)
method \cite{yip2003array}. Using the differences in the signal
intensity or energy level for the purpose of localization is a
more recent technique \cite{li2003energy,sheng2004maximum}.
Theoretically, this class of localization can be considered for
both narrowband and wideband signals by only taking into account
the attenuation information and usually neglecting the time delay
information. For these methods, a precise attenuation model in the
environment is inevitable for an accurate localization. Moreover,
from an optimization perspective the resulting cost functions in
these kind of approaches  usually undergo many local optima and
saddle points which require considering specific optimization
schemes \cite{blatt2006energy}.

In this paper, we tackle the problem of localization of multiple
wideband sources by coherently taking into account the TDOA and
the amplitude attenuation pattern. Our method generalizes the AML
approach to utilize the signal attenuation characteristics. We
provide a more robust algorithm in which the targets should
simultaneously satisfy the correct time delays among the sensors
and provide sensible level of attenuation at each sensor. Unlike
the aforementioned energy and intensity based methods which not
only ignore the time delay stamps but also require knowing the
signal attenuation model, we benefit using the delay information
and as a generalization to our recent work in
\cite{aghasiHashemiKhalaj}, leave the space for not knowing an
exact signal attenuation model in the environment by suggesting an
appropriate functional space for it. We minimize a cost function
which is obtained through maximum likelihood approach from which
the locations, attenuation model parameters and the multi-path
parameters are obtained. To apply the minimization we propose a
hybrid approach combining the differential evolution algorithm
\cite{storn1997differential} with the Levenberg-Marquardt
algorithm \cite{Madsen04methodsfor}. This combination provides a
minimization scheme which is likely to globally search for the
optima and rather quickly converges to the accurate results.
Through simulations and Cram\'{e}r-Rao bound we verify the
effectiveness of the novel method introduced in this paper.

This paper is organized as follows. In section \ref{sec2}, we
propose a general form for the received signal at every sensor and
later provide an adaptive model for the signal attenuation based
on Laurent polynomials. In section \ref{sec3}, a maximum
likelihood estimation of the source location and attenuation
parameters is proposed. We also provide the Cram\'{e}r-Rao bound
for this estimation problem. For the purpose of minimization in
section \ref{sec4} a hybrid approach combining the differential
evolution algorithm with the Levenberg-Marquardt is proposed for
which the combination algorithm and closed form equations for
calculation of the Jacobian are provided. In section \ref{sec5},
we examine the efficiency of proposed method through some examples
and finally there are some concluding remarks in section
\ref{sec6}.

\section{Problem Definition}\label{sec2}
\subsection{Signal Model}

Although the general approach proposed in this paper is in theory
independent of signal nature and the type of sensors used, in
order to make reasonable simulations we consider acoustic source
localization. Consider $N$ acoustic sources having unknown
locations $r_{S_n}.$ Each source is omni-directionally emitting a
signal $s_n(t),$ $n=1,\cdots,N$ at the time frame $t.$ We also
consider $M$ acoustic microphones that are placed in known
positions $r_{M_m},$ $m=1,\cdots,M$, in the same environment. For
every source in the environment, the function  that describes the
signal attenuation at a specific point is $\alpha(\rho),$ where
$\rho$ is the distance from the point to the source. In general,
the signal attenuation function may be a function of various
parameters such as signal frequency, medium inhomogeneity, etc. To
simplify the problem, in this paper we consider this function to
be an identical form for all sources and solely function of the
distance to the source. However, unlike some previous energy based
localizations (e.g., see \cite{sheng2004maximum, blatt2006energy})
in which the attenuation is known to be proportional to
$\rho^{-1}$, the actual form of $\alpha(\cdot)$ is considered
unknown function here. This type of modeling provides an
additional flexibility to the problem for more realistic scenarios
where the inverse proportionality of $\alpha(\cdot)$ to $\rho$ is
violated due to other parameters, such as signal bandwidth and
medium inhomogeneity. Considering $s_n(t-1\times N_s/v)$ to be the
signal measured 1 length unit away from every source with $N_s$
being the samples per second and $v$ being the propagation speed,
ideally the overall received signal samples from the acoustic
sources at every microphone is modeled as
\begin{align}
\label{eq1}
x_m(t)=\sum_{n=1}^{N}\alpha(\rho_{m,n})s_n(t-\tau_{m,n}),
\end{align}
for
\begin{equation*}
t=0,1,\cdots,n_t-1, \;\; m=1,2,\cdots,M.
\end{equation*}
Here $\rho_{m,n}=\|r_{M_m}- r_{S_n}\|$ is the distance from
$n^{\mbox{th}}$ source to $m^{\mbox{th}}$ microphone and
$\tau_{m,n}=\rho_{m,n}N_s/v$ is the corresponding time samples
delay in receiving the signal. The received signal in (\ref{eq1})
is normalized to each microphone gain level in order to decrease
the number of notations. A more realistic model takes into account
the noise in the environment and also the signals going through a
multi-path channel before arriving at every sensor, hence we
rewrite the received signal as
\begin{align}
\label{eq1-2}\nonumber
x_m(t)&=\sum_{n=1}^{N}\alpha(\rho_{m,n})s_n(t-\tau_{m,n})\\\nonumber
&+\sum_{n=1}^{N}\sum_{p=1}^{P_{m,n}}\gamma_{m,n,p}s_n(t-\hat\tau_{m,n,p})\\&+w_m(t).
\end{align}
The term $w_m(t)$ represents the background noise which is
considered to be a zero-mean white Gaussian with variance
$\sigma^2$ for the sake of this paper; i.e., Gaussianity is not a
limiting assumption in this paper. Between the $n^{\mbox{th}}$
source and $m^{\mbox{th}}$ microphone we consider $P_{m,n}$
indirect paths each causing $\gamma_{m,n,p}$ loss in the signal
amplitude and $\hat\tau_{m,n,p}$ delay in the signal reception,
modeling the multi-path phenomenon. Beside the positions
$r_{S_n},$ which are the main unknowns of the localization
problem, the signals $s_n(t)$, the multi-path parameters
$\gamma_{m,n,p}$ and $\hat\tau_{m,n,p}$, and the propagation loss
function $\alpha(\cdot)$ are also unknown and should be determined
based on the received signals at the sensors. The appearance of
$\tau_{m,n}$ (which is related to the unknown quantities
$r_{S_n}$) and $\hat\tau_{m,n,p}$ as the argument of an unknown
signal $s_n(t)$ causes an extra complexity for any optimization
scheme performed to solve the localization problem. However this
problem may be remedied by applying the discrete Fourier transform
to (\ref{eq1-2}) to extract the delays and form an equivalent
equation in which the unknown parameters are separated in
individual terms, i.e.,
\begin{align}
\label{eq2}  \nonumber
X_m(f)&=\sum_{n=1}^{N}\alpha(\rho_{m,n})\exp(-\frac{j2\pi}{n_f}f\tau_{m,n})S_n(f)\\\nonumber
&+\sum_{n=1}^{N}\sum_{p=1}^{P_{m,n}}\gamma_{m,n,p}\exp(-\frac{j2\pi}{n_f}f\hat\tau_{m,n,p})S_n(f)\\&+\xi_m(f)
\end{align}
for
\begin{equation*}
f=0,1,\cdots,n_f-1, \;\; m=1,2,\cdots,M.
\end{equation*}
Here, $X_m(f)$, $S_n(f)$ and $\xi_m(f)$ are the data, signal and
noise spectrums respectively. As stated in \cite{chen2002maximum},
we emphasize on the fact that for (\ref{eq2}) to be a valid
equivalent form of (\ref{eq1-2}), we need $n_t$ to be large enough
to avoid edge effects and accordingly $n_f>n_t$. In general having
more samples from the signal better poses the problem.

\subsection{A Low Order Representation of Signal Attenuation Model} As discussed earlier, our
assumption about the attenuation  model in the environment in this
paper is an identical model for all sources, which is only
dependent on the distance of the point to the acoustic source. In
an ideal environment, $\alpha(\rho)$ can be well modeled as a
multiple of $\rho^{-1}$. Since there are  different parameters
involved in the attenuation modeling, $\alpha(\rho)$ is being
considered  as an unknown here. However, in order to keep the
well-posedness of the problem, we choose it to be an element of a
low dimensional function space. For this sake, we consider
$\alpha(\rho)$ to be a \emph{Laurent polynomial} of limited order
as
\begin{equation}
\label{eq3} \alpha(\rho)=\sum_{\ell=1}^{L}\beta_l \rho^{-\ell},
\quad L>0.
\end{equation}
In this model, only negative powers of $\rho$ are considered,
which is due to the fact that for an attenuation model we are
physically required to have
\begin{equation}
\label{eq4} \lim_{\rho\rightarrow\infty}\alpha(\rho)=0.
\end{equation}
In many applications the low order representation of
$\alpha(\rho)$ in (\ref{eq3}) is acceptable enough to model the
attenuation  and usually considering only few terms in the series
(i.e., $L$ rather small), would suffice for the localization
problem.

\section{A Maximum Likelihood Estimation of
the Unknowns}\label{sec3}
\subsection{Derivation}
Based on the general attenuation model proposed, matching of the
data spectrum with the model can be expressed by using (\ref{eq3})
in (\ref{eq2}) as
\begin{small}
\begin{align}
\label{eq5}   \nonumber X_m(f)&=
\sum_{n=1}^{N}\sum_{\ell=1}^{L}\beta_\ell
\rho_{m,n}^{-\ell}\exp(-\frac{j2\pi}{n_f}f\tau_{m,n})S_n(f)\\\nonumber
&+\sum_{n=1}^{N}\sum_{p=1}^{P_{m,n}}\gamma_{m,n,p}\exp(-\frac{j2\pi}{n_f}f\hat\tau_{m,n,p})S_n(f)\\&+\xi_m(f).
\end{align}
\end{small}
The central limit theorem states that $\xi_m(f)$, which is a
transformed zero mean Gaussian random variable to the frequency
domain, should be a complex zero mean Gaussian with variance
$n_t\sigma^2$. For every frequency bin $f$ having
$\boldsymbol{X}(f)=[X_1(f),\cdots,X_{M}(f)]^T$,
$\boldsymbol{S}(f)=[S_1(f),\cdots,S_{N}(f)]^T$ and
$\boldsymbol{\xi}(f)=[\xi_1(f),\cdots,\xi_{M}(f)]^T$, (\ref{eq5})
can be written in a matrix form as
\begin{equation}
\label{eq6}
\boldsymbol{X}(f)=\big(\boldsymbol{K}(f)+\boldsymbol{H}(f)\big)\boldsymbol{S}(f)+\boldsymbol{\xi}(f)
\end{equation}
where $\boldsymbol{K}(f)=\boldsymbol{R}(f)\boldsymbol{\beta}$ with
\begin{equation}
\boldsymbol{\beta}=[\beta_1,\cdots,\beta_{L}]^T\otimes
\boldsymbol{I}_{N\times N},
\end{equation}
for which $\otimes$ represents the Kronecker product and
$\boldsymbol{I}_{N\times N}$ the identity matrix of size $N\times
N$, and
\begin{equation}
\boldsymbol{R}(f)=[\boldsymbol{R}_1(f),\cdots,\boldsymbol{R}_{L}(f)],
\end{equation}
where
\[\boldsymbol{R}_\ell(f)=\left [
\begin{array}{ccc}
\hspace{-.2cm}\rho_{1,1}^{-\ell}e^{-\frac{j2\pi
N_s}{n_fv}f\rho_{1,1}}&\hspace{-.2cm}\cdots\hspace{-.2cm}
&\rho_{1,N}^{-\ell}e^{-\frac{j2\pi N_s}{n_fv}f\rho_{1,N}}\\
\vdots& \hspace{-.2cm}\ddots\hspace{-.2cm}& \vdots
\\\hspace{-.2cm} \rho_{M,1}^{-\ell}e^{-\frac{j2\pi
N_s}{n_fv}f\rho_{M,1}}&\hspace{-.2cm}\cdots\hspace{-.2cm}
&\rho_{M,N}^{-\ell}e^{-\frac{j2\pi N_s}{n_fv}f\rho_{M,N}}\\
\end{array}\hspace{-.2cm} \right ],
\]
for $\ell=1,\cdots, L.$ The matrix $\boldsymbol{H}(f)$ is related
to the multi-path parameters and its elements are
\begin{equation}
[\boldsymbol{H}(f)]_{(m,n)}=\sum_{p=1}^{P_{m,n}}\gamma_{m,n,p}\exp(-\frac{j2\pi}{n_f}f\hat\tau_{m,n,p}).
\end{equation}
Rewriting the  negative log-likelihood function to estimate the
unknown parameters ${\theta}$ including the source positions,
source signal spectrums, multi-path parameters and quantities
$\beta_\ell,$ we have
\begin{equation}
\label{eq9} {\theta}^*=\mbox{arg} \min_{{\theta}}
\boldsymbol{Q}^H\boldsymbol{Q}
\end{equation}
where
\begin{equation}\label{eq10}
\boldsymbol{Q}= \left [
\begin{array}{c}
\boldsymbol{Q}(0)\\ \vdots \\ \boldsymbol{Q}({n_f}/{2})
\end{array}\right ]
\end{equation}
and $\boldsymbol{Q}(f)=\boldsymbol{X}(f)-\boldsymbol{\tilde
K}(f)\boldsymbol{S}(f)$ using the notation
\begin{equation}\label{eq10b}
\boldsymbol{\tilde K}(f)=\boldsymbol{ K}(f)+\boldsymbol{H}(f).
\end{equation}
Similar to the idea in \cite{chen2002maximum}, for a real valued
signal, we can only consider up to $n_f/2$ frequency bins and form
$\boldsymbol{Q}$ with blocks of $\boldsymbol{Q}(f)$ for
$f=0,1,\cdots,n_f/2$. We would like to highlight the fact that in
\cite{chen2002maximum}, the zero frequency bin is ignored due to
producing a constant term in the likelihood function, however in
our approach the matrices $\boldsymbol{K}(0)$ and
$\boldsymbol{H}(0)$ are still dependent on $\rho_{m,n}$ and the
multi-path parameters $\gamma_{m,n,p}$ and hence worth being
considered.

Clearly, the minimization in (\ref{eq9}) is equivalent to
minimizing $\boldsymbol{Q}^H(f)\boldsymbol{Q}(f)$ for every $f$.
Considering the unknown signal spectrum $\boldsymbol{S}(f)$, the
minima should satisfy
\begin{equation}
\frac{\partial
\big(\boldsymbol{Q}^H(f)\boldsymbol{Q}(f)\big)}{\partial
\boldsymbol{S}^H(f)}=0
\end{equation}
which results in $\boldsymbol{S}(f)=\boldsymbol{\tilde
K}^\dag(f)\boldsymbol{X}(f)$ with $\boldsymbol{\tilde K}^\dag(f)$
representing the pseudo-inverse of $\boldsymbol{\tilde K}(f)$.
Replacing the obtained $\boldsymbol{S}(f)$ in $\boldsymbol{Q}(f)$
results in
\begin{align}
\label{eq12}
\boldsymbol{Q}(f)=\boldsymbol{X}(f)-\boldsymbol{\tilde
K}(f)\boldsymbol{\tilde K}^\dag(f)\boldsymbol{X}(f),
\end{align}
for $f=0,\cdots,n_f/2$, and therefore the unknowns of the
minimization reduce to the source positions, multi-path parameters
and the attenuation coefficients. Considering a 2D localization
problem, as the case in the example section, neglecting the
multi-path the vector of unknowns would be
\begin{equation}\label{eqtheta}
\boldsymbol{\theta}=[x_{S_1},\cdots,x_{S_{N}},y_{S_1},\cdots,y_{S_{N}},\beta_1,\cdots,\beta_{L}]^T,
\end{equation}
where $x_{S_n}$ and $y_{S_n}$ are the $x$ and $y$ components of
the position vector $r_{S_n}$. In case of multi-path, the
parameters $\gamma_{m,n,p}$ and $\hat\tau_{m,n,p}$ are also
included in $\boldsymbol{\theta}$. The approach is clearly not
only limited to 2D Cartesian systems and 3D Scenarios and other
coordinate systems may be considered.

\subsection{Cram\'{e}r-Rao Lower Bounds for the Estimated Parameters} \label{secCRLB}
For an unbiased parameter estimation problem, the Cram\'{e}r-Rao
Lower Bound (CLRB) is a theoretical lower bound on the variance of
the problem estimates. Based on (\ref{eq6}), the total model
relating the parameters of interest to the complete data set is
\begin{equation}
\label{eqmodel}
\boldsymbol{X}=\boldsymbol{\mathcal{G}}(\boldsymbol{\theta};\boldsymbol{S})+\boldsymbol{\xi}.
\end{equation}
Here
$\boldsymbol{X}=[\boldsymbol{X}(0)^T,\cdots,\boldsymbol{X}(n_f/2)^T]^T$
is the full data set,
$\boldsymbol{S}=[\boldsymbol{S}(0)^T,\cdots,\boldsymbol{S}(n_f/2)^T]^T$
contains the signal spectrum of all the sources and
$\boldsymbol{\xi}=[\boldsymbol{\xi}(0)^T,\cdots,\boldsymbol{\xi}(n_f/2)^T]^T$
is the corresponding noise vector. Moreover,
$\boldsymbol{\mathcal{G}}(\boldsymbol{\theta};\boldsymbol{S})=\boldsymbol{\tilde
K}\boldsymbol{S}$ for which the matrix $\boldsymbol{\tilde K}$
explicitly dependent on $\boldsymbol{\theta}$ is
\begin{equation}
\boldsymbol{\tilde K}= \left [
\begin{array}{cccc}
\boldsymbol{\tilde K}(0)&\boldsymbol{0}&\cdots&\boldsymbol{0}\\
\boldsymbol{0}&\boldsymbol{\tilde K}(1)&\cdots&\boldsymbol{0} \\
\vdots& \vdots& \ddots &\vdots\\
\boldsymbol{0}&\boldsymbol{0}&\cdots&\boldsymbol{\tilde
K}({n_f}/{2})
\end{array}\right ].
\end{equation}
The CRLB is defined as the diagonal elements of the inverse Fisher
matrix $\boldsymbol{F}$, which for the model in (\ref{eqmodel}) is
representable as \cite{kay1993fundamentals}
\begin{equation}
\boldsymbol{F}=\Big[\frac{\partial
\boldsymbol{\mathcal{G}}}{\partial \boldsymbol{\vartheta}}\Big]^H
\boldsymbol{\mathcal{R}}_{\xi}^{-1}\Big[\frac{\partial
\boldsymbol{\mathcal{G}}}{\partial \boldsymbol{\vartheta}}\Big].
\end{equation}
Here
\begin{equation}
\boldsymbol{\vartheta}=\left [
\begin{array}{c}
\boldsymbol{S}\\ \boldsymbol{\theta}
\end{array}
\right ],
\end{equation}
and $\boldsymbol{\mathcal{R}}_{\xi}$ is the noise covariance
matrix which for our problem is simply
$n_t\sigma^2\boldsymbol{I}$. The matrix $\big[{\partial
\boldsymbol{\mathcal{G}}}/\partial \boldsymbol{\vartheta}\big ]$
is composed of the sub-blocks $\big[{\partial
\boldsymbol{\mathcal{G}}}/\partial \boldsymbol{S}\big ]$,
$\big[{\partial \boldsymbol{\mathcal{G}}}/\partial r_{S_n} ]$,
$\big[{\partial \boldsymbol{\mathcal{G}}}/\partial
\beta_\ell\big]$, $\big[{\partial
\boldsymbol{\mathcal{G}}}/\partial \gamma_{m,n,p} ]$ and
$\big[{\partial \boldsymbol{\mathcal{G}}}/\partial
\hat\tau_{m,n,p} ]$. Clearly
\begin{equation}
\frac{\partial \boldsymbol{\mathcal{G}}}{\partial
\boldsymbol{S}}=\boldsymbol{\tilde K}.
\end{equation}
For the $\boldsymbol{\theta}$ parameters, since
$\boldsymbol{\tilde K}$ is composed of $\boldsymbol{\tilde K}(f)$,
we only discuss the sensitivity of $\boldsymbol{\tilde K}(f)$ to
every class of parameters. Based on the fact that
$\boldsymbol{K}(f)=\boldsymbol{R}(f)\boldsymbol{\beta}$ we can
write
\begin{equation}\label{eqa6}
\frac{\partial \boldsymbol{\tilde K}(f)}{\partial
\beta_\ell}=\frac{\partial \boldsymbol{K}(f)}{\partial
\beta_\ell}=\boldsymbol{R}(f) \frac{\partial
\boldsymbol{\beta}}{\partial \beta_\ell},\quad \ell=1,2,\cdots,L
\end{equation}
where
\begin{equation}\label{eqa7}
\frac{\partial \boldsymbol{\beta}}{\partial \beta_\ell}=[0,\cdots,
0,\!\!\!\!\!\!\!\!\!\overbrace{1}^{{\mbox{\tiny
$\ell$}}^{\mbox{\tiny th}}\mbox{ \tiny
element}}\!\!\!\!\!\!\!\!\!,0,\cdots,0]^T\otimes
\boldsymbol{I}_{N\times N}.
\end{equation}
To calculate $\partial \boldsymbol{\tilde K}(f)/\partial x_{S_n}$,
we have
\begin{equation}\label{eqa8}
\frac{\partial \boldsymbol{\tilde K}(f)}{\partial
x_{S_n}}=\frac{\partial \boldsymbol{R}(f)}{\partial x_{S_n}}
\boldsymbol{\beta},\quad n=1,2,\cdots,N
\end{equation}
where
\begin{equation}\label{eqa9}
\frac{\partial \boldsymbol{R}(f)}{\partial
x_{S_n}}=\big[\frac{\partial \boldsymbol{R}_1(f)}{\partial
x_{S_n}},\cdots,\frac{\partial \boldsymbol{R}_{L}(f)}{\partial
x_{S_n}} \big].
\end{equation}
The matrix $\partial \boldsymbol{R}_\ell(f)/\partial x_{S_n}$ is a
matrix the same size as $\boldsymbol{R}_\ell(f)$, with all columns
being zero except the $n^{\mbox{th}}$ column. Simply applying the
derivative shows that the $(m,n)$ element of $\partial
\boldsymbol{R}_\ell(f)/\partial x_{S_n}$ is related to the $(m,n)$
element of $\boldsymbol{R}_\ell(f)$ through
\begin{small}
\begin{align}\nonumber
\Big[\frac{\partial \boldsymbol{R}_l(f)}{\partial
x_{S_{n'}}}\Big]_{m,n}&=\delta(n,n')\left(\frac{x_{M_m}\!\!-\!x_{S_n}}{\rho_{m,n}}\right)\\
&\times\left(\frac{\ell}{\rho_{m,n}}+ \frac{j2\pi N_s
f}{n_fv}\right)\big[ \boldsymbol{R}_\ell(f)\big]_{m,n},
\end{align}
\end{small}
where
 \begin{displaymath}
   \delta(n,n') = \left\{
     \begin{array}{lr}
       1 ,&  n=n'\\
       0 ,&  n\neq n'
     \end{array}.
   \right.
\end{displaymath}
An analogous technique is used to derive $\partial
\boldsymbol{\tilde K}(f)/\partial y_{S_n}$.

For the multi-path parameters we have ${\partial
\boldsymbol{\tilde K}(f)}/{\partial \gamma_{m,n,p}}={\partial
\boldsymbol{H}(f)}/{\partial \gamma_{m,n,p}}$ and also have
${\partial \boldsymbol{\tilde K}(f)}/{\partial
\hat\tau_{m,n,p}}={\partial \boldsymbol{H}(f)}/{\partial
\hat\tau_{m,n,p}}$. Accordingly the elements of each matrix are
obtained through
\begin{equation}
\Big[\frac{\partial \boldsymbol{H}(f)}{\partial
\gamma_{m',n',p}}\Big]_{(m,n)}\!\!\!=\delta(m,m')\delta(n,n')\exp(-\frac{j2\pi}{n_f}f\hat\tau_{m,n,p})
\end{equation}
and
\begin{align}\nonumber
\Big[\frac{\partial \boldsymbol{H}(f)}{\partial
\hat\tau_{m',n',p}}\Big]_{(m,n)}&=-\delta(m,m')\delta(n,n')\frac{j2\pi}{n_f}f\gamma_{m,n,p}\\&\times\exp(-\frac{j2\pi}{n_f}f\hat\tau_{m,n,p}).
\end{align}
Specifying the elements of the Fisher matrix $\boldsymbol{F}$
yields the CRLB values for all the estimations.

\section{Minimization Strategy}\label{sec4}
The minimization in (\ref{eq9}) may be performed through various
optimization schemes, most generally categorized as \emph{global}
and \emph{local} optimizations. For a global optimization
different approaches such as deterministic, stochastic or
evolutionary and metaheuristic methods may be considered
\cite{horst1996global, spall2003introduction, glover2003handbook}.
Clearly for an accurate localization, global minimizers of
(\ref{eq9}) are required. However in general, using global methods
to optimize an arbitrary function may be iteratively or
computationally expensive. As an alternative to this and
specifically for a least squares cost function as  (\ref{eq9}),
local search methods such as gradient descent and quasi-Newton
methods may be considered \cite{Madsen04methodsfor}. Although
these methods can be relatively faster than the global ones, there
is always a chance of getting trapped into a local minima. In the
context of localization, although for good initial estimates of
the source relatively fast methods such as the gradient descent
and alternating projection are proposed, to increase the chances
of finding a global minima the process usually involves exhaustive
search methods such as the grid search and multiresolution search
\cite{chen2002maximum, sheng2004maximum}.

For the purpose of this paper we consider a hybrid approach
combining a global search method with a fast local search method
\cite{coelho2006combining, preux1999towards}. Hybrid methods have
received considerable interests in different areas in the recent
years
\cite{bube1997hybrid,esmin2005hybrid,fujita1993hybrid,preux1999towards}.
More specifically we consider a hybrid combination of the
Differential Evolution algorithm (DE) \cite{storn1997differential}
as successful evolutionary search with the Levenberg-Marquardt
algorithm (LMA) \cite{Madsen04methodsfor,dennis1996numerical} as a
rather fast and robust local search method. Before getting to the
combination scheme, we provide a brief description of each method
highlighting the main technical issues specifically in the context
of our localization problem.
\subsection{Differential Evolution Algorithm}
DE is among the metaheuristic and evolutionary global optimization
schemes. Simplicity and successful performance are the main
advantages of this algorithm. Considering
$\boldsymbol{\theta}=[\theta_1,\theta_2,\cdots ,\theta_D]$ to be
the vector of problem unknowns with size $D$, at every generation
$G$ of the algorithm $N_P$ parameter vectors
$\boldsymbol{\theta}_{i,G}=[\theta_{1,i,G},\theta_{2,i,G},\cdots
,\theta_{D,i,G}]$, $i=1,2,\cdots,N_P,$ are generated. The initial
population is randomly chosen with a uniform distribution in the
search region. For this work we consider the $DE/rand/1/bin$,
which is a general and widely used strategy of this algorithm
\cite{storn1997differential, brest2006self}. For every generation
three main operations are performed as follows.
\subsubsection{Mutation} In this phase a mutant vector
$\boldsymbol{\mu}_{i,G}$ is generated as
\begin{equation}
\boldsymbol{\mu}_{i,G}=\boldsymbol{\theta}_{r_1,G}+F\;(\boldsymbol{\theta}_{r_2,G}-\boldsymbol{\theta}_{r_3,G}),
\end{equation}
where $r_1$, $r_2$ and $r_3$ are randomly selected indices among
$1,2,\cdots,N_P$ and $F\in [0,2]$ is a constant real scalar
controlling the difference vector amplification.
\subsubsection{Crossover} A mixing with the mutant vector to
increase the diversity of the population is performed by
generating new trial vectors $\boldsymbol{\upsilon}_{i,G}$ of
length $D$, defined as
\begin{equation}
   \upsilon_{d,i,G} = \left\{
     \begin{array}{ll}
       \mu_{d,i,G}, & \quad r(d)_{[0,1]}\leq C_R\;\mbox{or}\;d=k(i)\\
       \theta_{d,i,G}, & \quad \mbox{otherwise,}
     \end{array}
   \right.
\end{equation}
with $d=1,2,\cdots,D$. Here $C_R\in [0,1]$ is the crossover
constant, $r(d)_{[0,1]}$ is the $d^{\mbox{th}}$ evaluation of a
uniform random number generator in $[0,1]$ and $k(i)\in
\{1,2,\cdots,D\}$ is a randomly chosen index ensuring that
$\boldsymbol{\upsilon}_{i,G}$ takes at least one of the elements
of $\boldsymbol{\mu}_{i,G}$.
\subsubsection{Selection} At this step a next generation population member $\boldsymbol{\theta}_{i,G+1}$
is produced by a selection among $\boldsymbol{\theta}_{i,G}$ and
$\boldsymbol{\upsilon}_{i,G}$. This selection is based on the
fitness, and basically, the vector with the lower cost proceeds to
the next generation.
\subsection{A Levenberg-Marquardt Algorithm for the local
Minimization} As the local minimization scheme, we suggest using
the LMA. Our attention towards this algorithm is based on several
advantages. LMA is basically considered as a Newton type method
and provides a rather quadratic convergence. Meanwhile this
algorithm benefits from stability and uses a trust region approach
\cite{dennis1996numerical}. The other feature of this method,
considered as an advantage over other methods such as the gradient
descent, is its suitability for cases where there are different
variables of different types as the cost function arguments. In
fact LMA is almost independent of variable scaling, while for
methods such as the gradient descent, minimizing a cost function
dependent on a set of variables with different natures and scales
requires appropriate parameter scaling to guarantee a proper
convergence \cite{dennis1996numerical}. This is a demanding
feature for our problem where the $\boldsymbol{\theta}$ vector in
general consists of the source locations, attenuations
coefficients and the multi-path parameters.

In the LMA which is an iterative algorithm, we start with a
$\boldsymbol{\theta}^{(0)}$ as the starting point. At every
iteration, having $\boldsymbol{\theta}^{(i)}$ already in hand,
$\boldsymbol{\theta}^{(i+1)}$ can be obtained by solving
\begin{equation}
(\boldsymbol{J_\theta}^T\boldsymbol{J_\theta}+\lambda^{(i)}\boldsymbol{I})(\boldsymbol{\theta}^{(i+1)}-\boldsymbol{\theta}^{(i)})=-\boldsymbol{J_\theta}^T\boldsymbol{Q},
\end{equation}
where $\boldsymbol{Q}$ is the vertical vector of length $Mn_f/2$
shown in (\ref{eq10}) and obtained for values
$\boldsymbol{\theta}^{(i)}$ at that iteration. The parameter
$\lambda^{(i)}$ is the damping factor, obtained at every iteration
based on the trust region approach
\cite{Madsen04methodsfor,dennis1996numerical}. The Jacobian matrix
$\boldsymbol{J_\theta}$ contains the sensitivities of
$\boldsymbol{Q}$ to every element of $\boldsymbol{\theta}$. In
order to run the algorithm we need to know the Jacobian matrix at
every iteration, obtaining which is discussed in the Appendix.

\subsection{The Hybrid Combination Scheme}
For the purpose of combining the DE with the LMA, we propose using
a \emph{sequential hybridization} approach
\cite{preux1999towards}. In this approach, the DE initially starts
the minimization by generating consecutively more fitting
generations. After a certain number of generations or after
getting relatively slow in decreasing the fitness, the best
$\boldsymbol{\theta}$ in the last generation is passed to the LMA
algorithm as an initialization. The minimization continues until
convergence. An illustration of this algorithm is provided in Fig.
\ref{algorithm}.
\begin{figure*}\centering
\epsfig{file=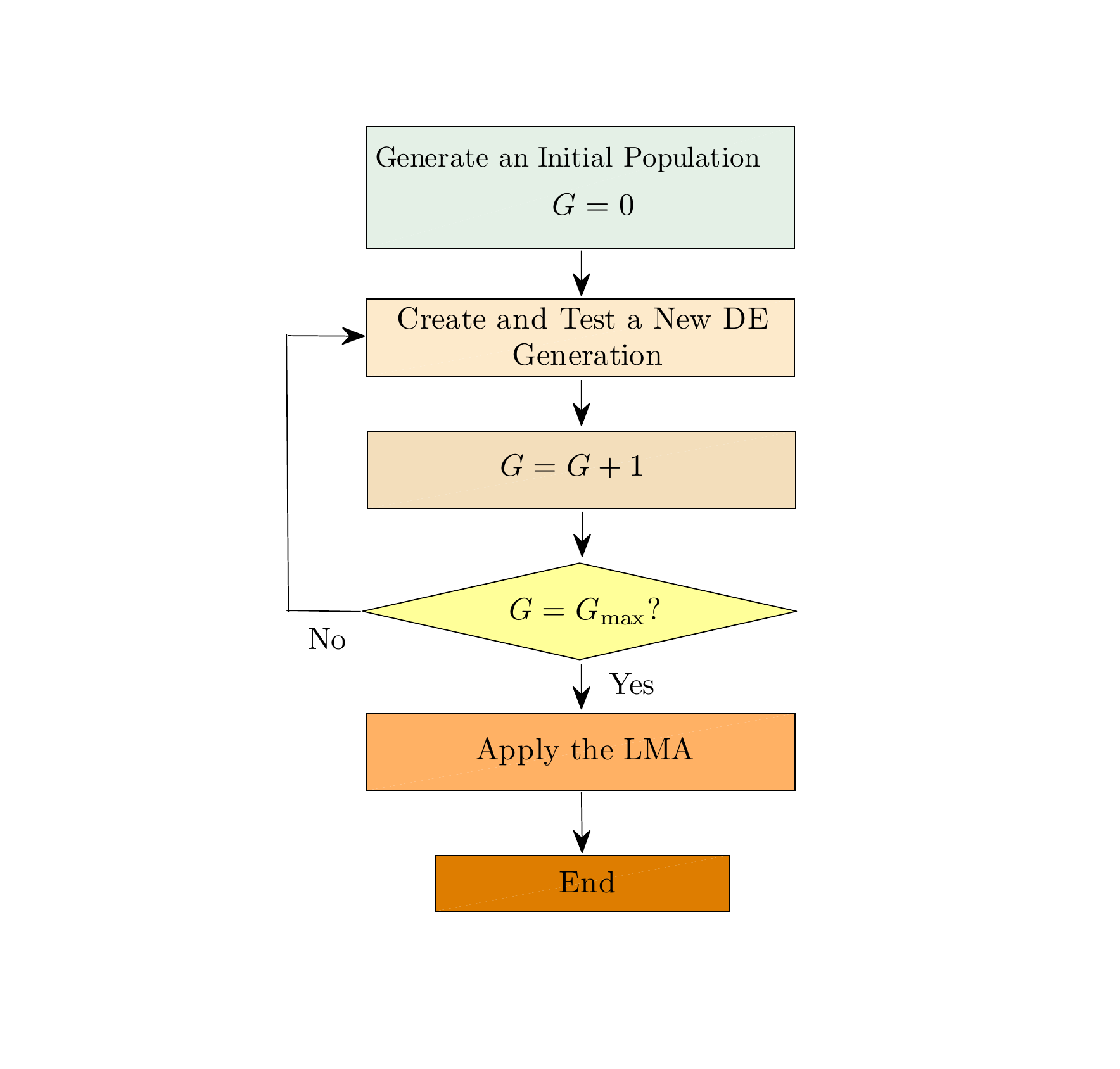,width=.55\linewidth,clip=} \caption{A
sequential hybridization combing the DE with the
LMA}\label{algorithm}
\end{figure*}

\section{Simulation Results} \label{sec5}
To examine the method developed in the previous section, we
consider some localization examples in this section. In the first
example we consider a reverberation-free environment to show the
efficiency of the method for such cases and provide a comparative
study for this scenario. The second example brings more realistic
issues such as the multi-path, and sensor synchronization error
into the problem and examines the performance of the proposed
method for such cases.

Before proceeding with the examples we would like to highlight a
fact regarding the relationship between the cost function and the
matrix $\boldsymbol{\tilde K}$. Referring to (\ref{eq12}), it can
be easily verified that scaling $\boldsymbol{\tilde K}(f)$ by a
scalar does not change the cost function. In other words if the
$\beta_\ell$ and $\gamma_{m,n,p}$ values are simultaneously scaled
by a scalar value, the cost function remains the same. Therefore,
we rewrite the attenuation model in (\ref{eq3}) as
\begin{equation}
\label{eq14} \alpha(\rho)={\rho}^{-1}+\sum_{\ell=1}^{L}\beta_\ell
\rho^{-\ell-1},
\end{equation}
which somehow normalizes $\alpha(\cdot)$ and unifies the
representation. Clearly, since the desired unknowns of the problem
are the acoustic source coordinates, obtaining a multiple of the
attenuation and multi-path coefficients is non-problematic. The
true attenuation model to be used in this paper is
$\alpha(\rho)=\rho^{-1.25}$ (see \cite{li2003energy}).

\subsection{Example 1}
For the purpose of this example, we consider the sensors to be
placed in the first quarter of the x-y plane as a spiral array of
$M=40$ microphones. The spiral is represented in a parametric form
as
\begin{equation}
\left [
\begin{array}{c}
x_{M_m}\\y_{M_m}
\end{array}
\right ] = \left [
\begin{array}{c}
4+\frac{s}{\pi}\cos{s}\\4+\frac{s}{\pi}\sin{s}
\end{array}
\right ],
\end{equation}
where the angles $s$ are equally spaced in $[2\pi,4\pi]$. Our
purpose of choosing such sensor arrangement was to provide a
non-symmetric and still reproducible arrangement. The sensor
locations are shown in Fig. \ref{fig2}. The sources used in this
example are wideband sources with center frequency 500 Hz and 200
Hz bandwidth. The sampling frequency is 4 KHz. The number of
samples available from the sources at every sensor is $n_t=4000$
(i.e., the signal duration is 1 second) and the number of
frequency bins is taken to be $n_f=4100$. The signal to noise
ratio (SNR) at every sensor is 20 dB and the speed of propagation
is considered to be the speed of sound as $v=345\;m/s$. In the
proposed minimization scheme and more specifically the DE part, we
take $G_{\max}=5$. Moreover, we set $F=0.8$, $C_R=1$ and $N_P=40$.
This parameter setting was selected as a general DE setting,
however more discussions on determining the DE parameters are
available in \cite{storn1997differential}. The general attenuation
model is considered to be
$\alpha(\rho)=\rho^{-1}+\beta_1\rho^{-2}+\beta_2\rho^{-3}$, for
which the values $\beta_1$ and $\beta_2$ are in charge of tuning
the unknown model. There is no reverberation in the environment
(i.e., $\boldsymbol{\tilde K}=\boldsymbol{K}$) and all sensors are
synchronized in receiving the signal.
\begin{figure*}
\hspace{-.4cm}{\subfigure[ ]{\includegraphics[width=3.3in
]{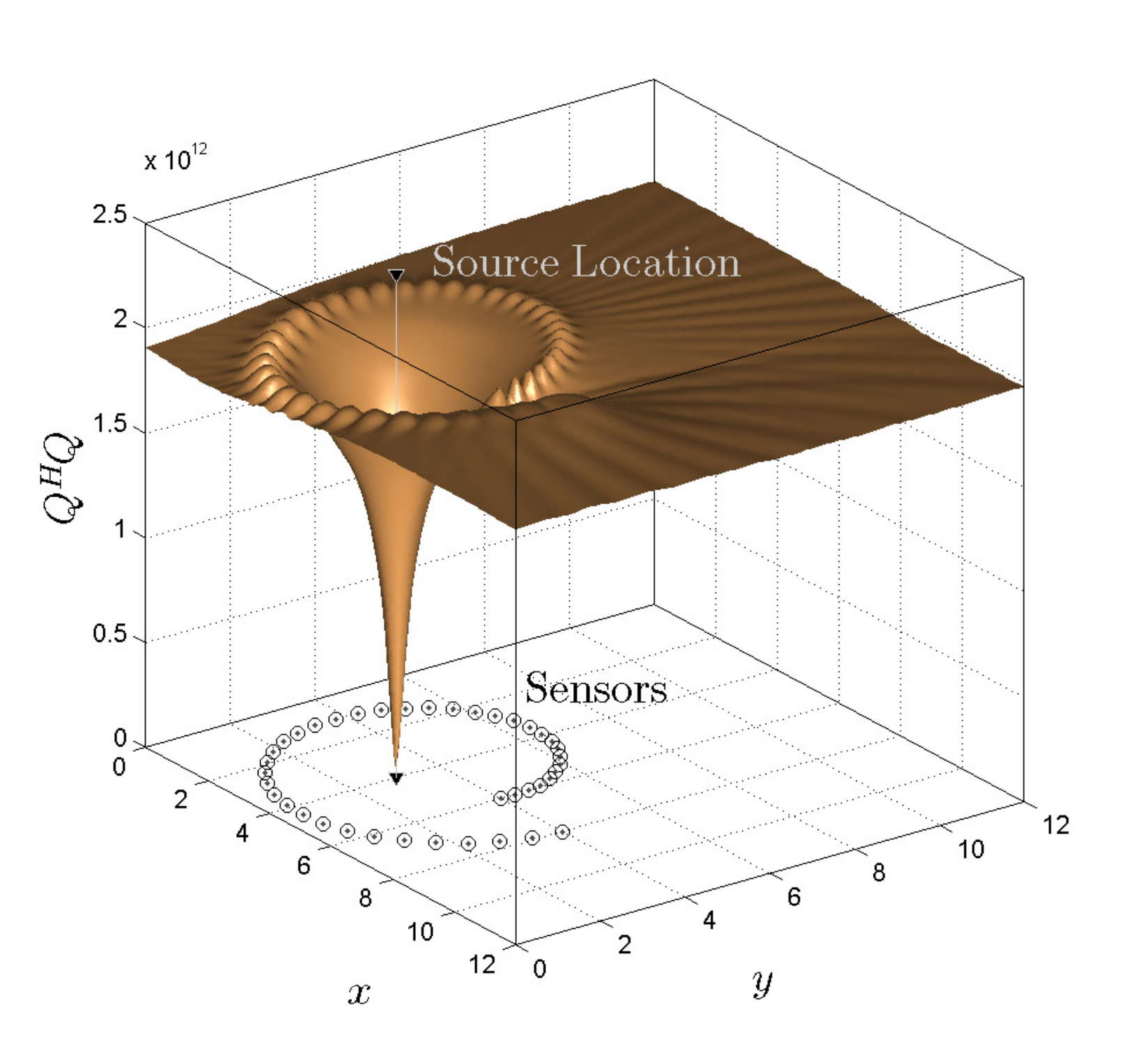}}\hspace{-.4cm} \subfigure[ ]{\includegraphics[width=3.3in
]{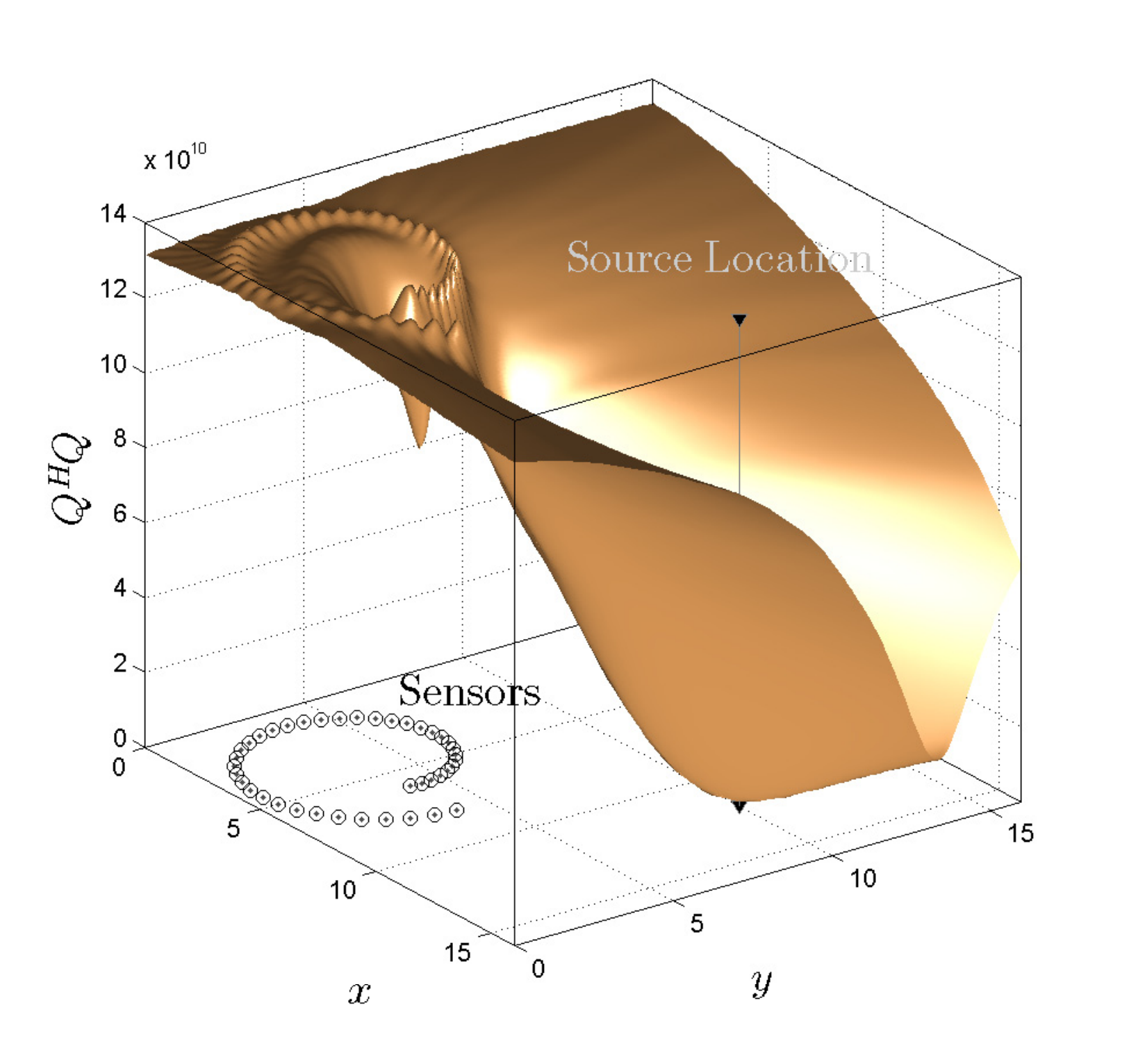} }\hspace{-.5cm}} \caption{(a) The cost function
corresponding to a source located at $(4,3)$ assuming a known
attenuation model. (b) The cost function corresponding to a source
located at $(12,10)$.} \label{fig2}
\end{figure*}

To provide a better understanding of the problem, in Fig.
\ref{fig2} the cost function behavior for a known attenuation
model is shown. In Fig. \ref{fig2}.a the cost is shown when the
source is located at point $(4,3)$ within the sensors convex hull.
All positions are in meters. Fig. \ref{fig2}.b shows the cost when
the source is located at $(12,10)$ outside the sensors region. In
both cases the cost functions are rather well behaved functions
away from the sensors. Intuitively, for two neighboring points in
the domain, sudden variation of the cost function with respect to
both time delay criteria and attenuation model constraints is
unlikely and hence the resulting cost functions are usually
expected to be rather slow varying and well behaved away from the
sensors.

In Fig. \ref{fig3} we have shown the iterative procedure of
finding a single source, once located at $(4,3)$ and once at
$(12,10)$. For the first case the source location is estimated to
be at $(4.002,3.008)$ and the attenuation coefficients are
estimated to be $\beta_1=-23.85$ and $\beta_2=27.93$. In the
second case the source estimation is $(11.999,10.002)$ and the
attenuation coefficients are found to be $\beta_1=4.19$ and
$\beta_2=1.79$. We observe  that both localization results
accurately match the exact source positions. The attenuation
coefficients obtained in both cases are only in charge of fitting
the low order model to the true model for the source-sensor ranges
in each problem and due to different problems they do not
necessarily need to be in the same ranges. By providing this low
order attenuation model we provide the flexibility to the problem
for accurately estimating the sources.
\begin{figure*}
\hspace{-.5cm}{\subfigure[ ]{\includegraphics[width=2.25in
]{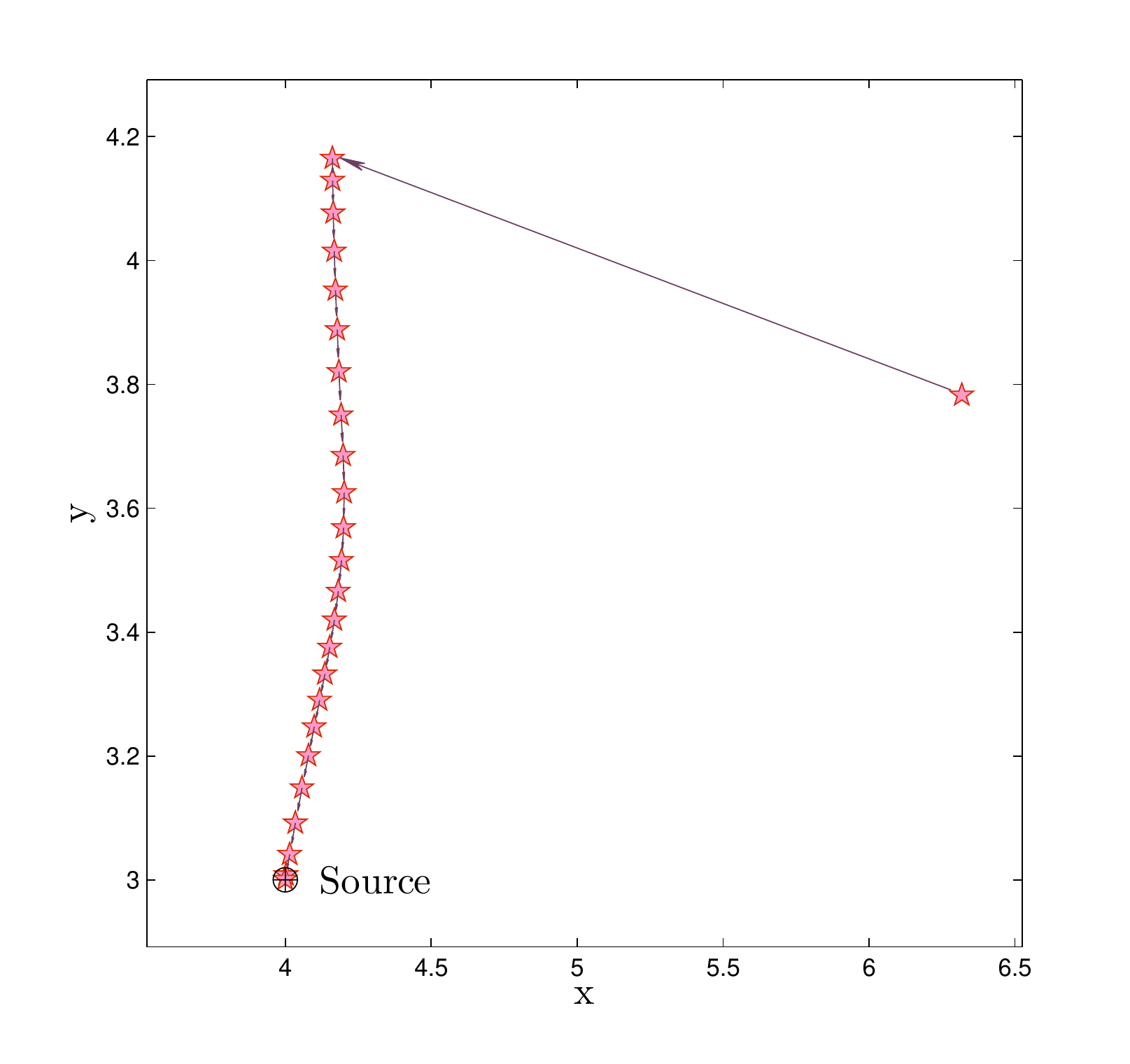}} \subfigure[ ]{\includegraphics[width=2.25in ]{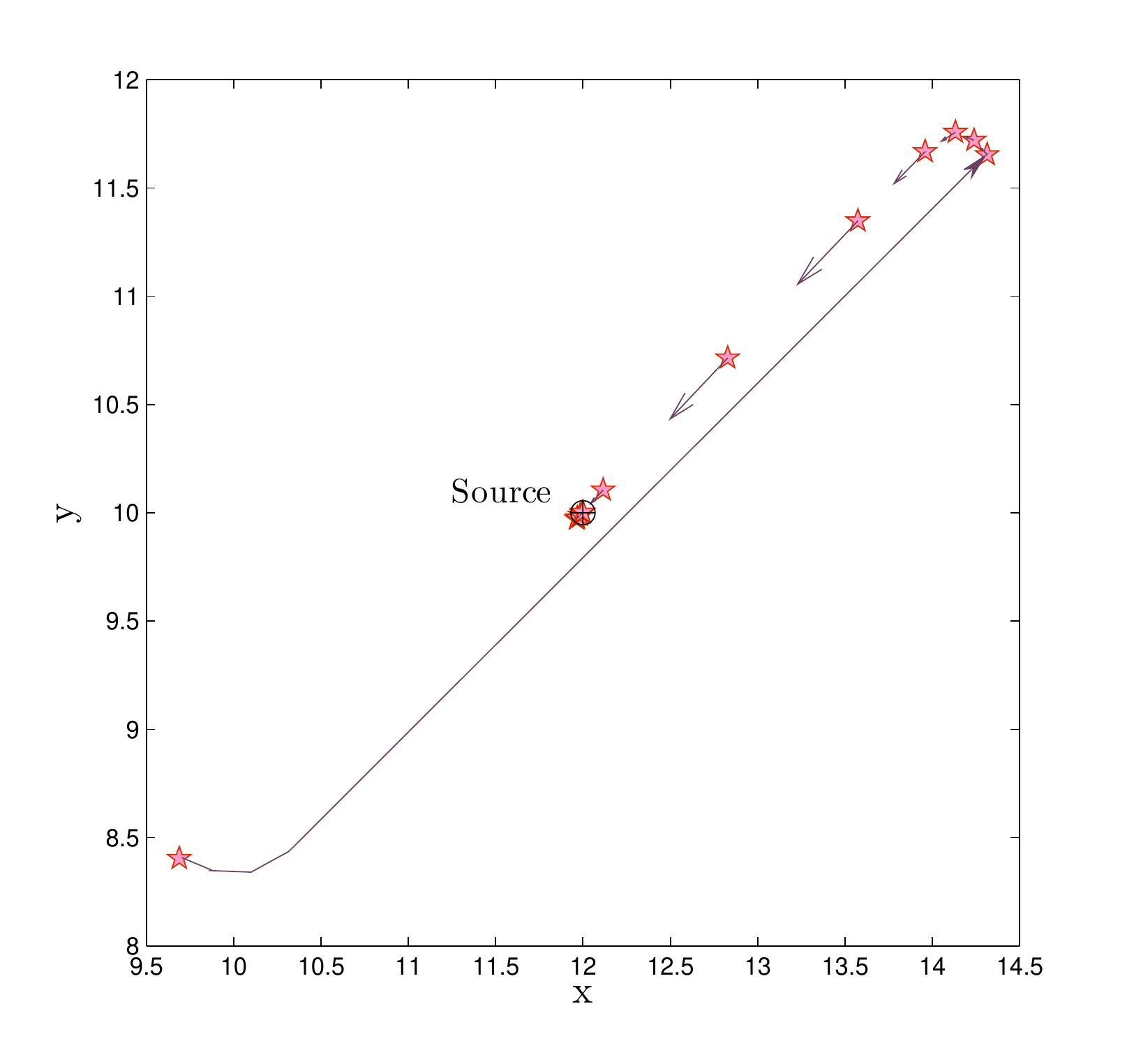}
}\subfigure[ ]{\includegraphics[width=2.25in ]{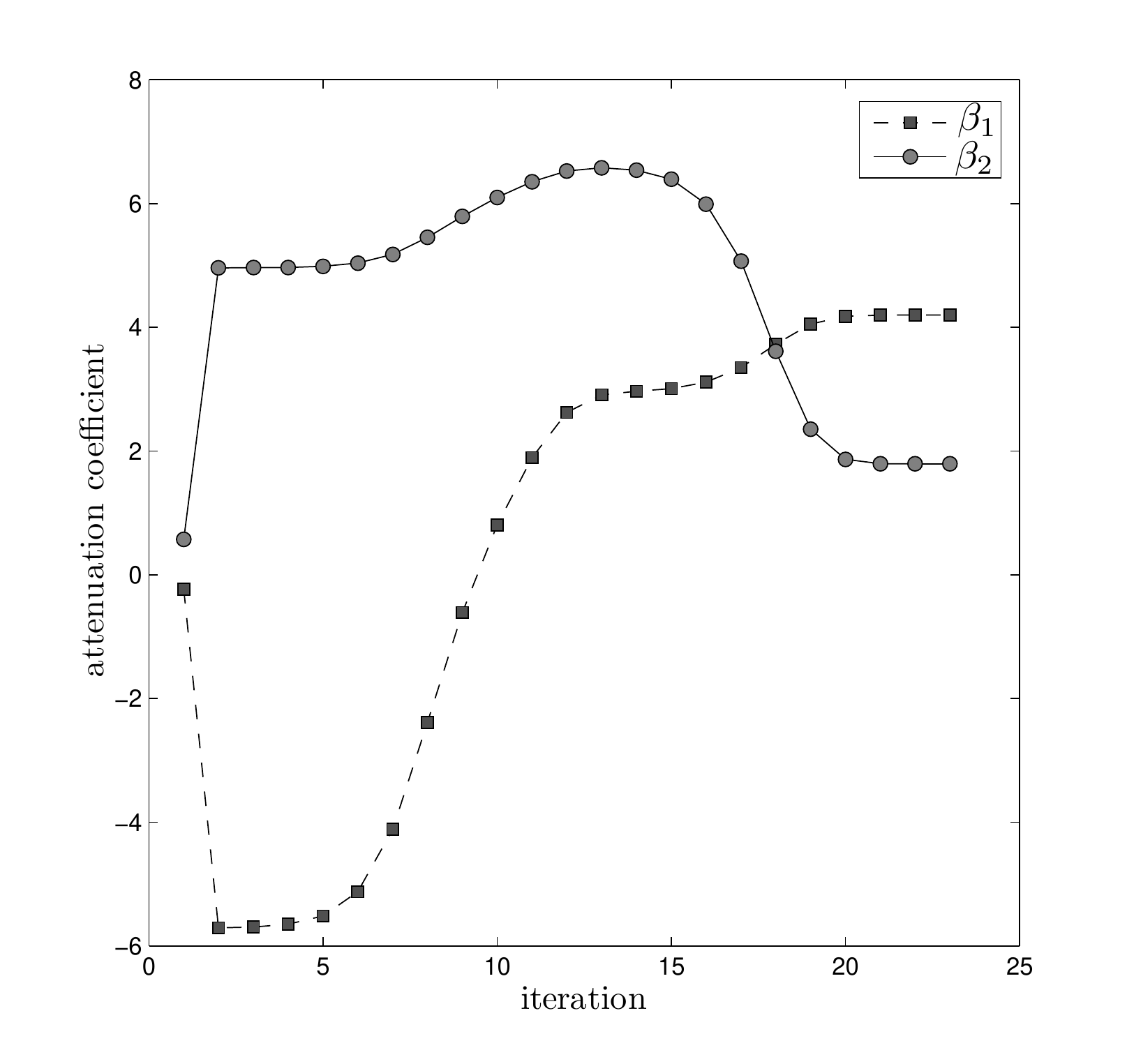} }}
\caption{(a) The progressive estimates of a single source located
at $(4,3)$. The first jumps and good initials correspond to
applying the DE (b) The progressive estimates of a single source
located at $(12,10)$. (c) The evolution of the attenuation model
parameters for the single source located at $(12,10)$}
\label{fig3}
\end{figure*}

\begin{figure*}
{\subfigure[ ]{\includegraphics[width=3.2in ]{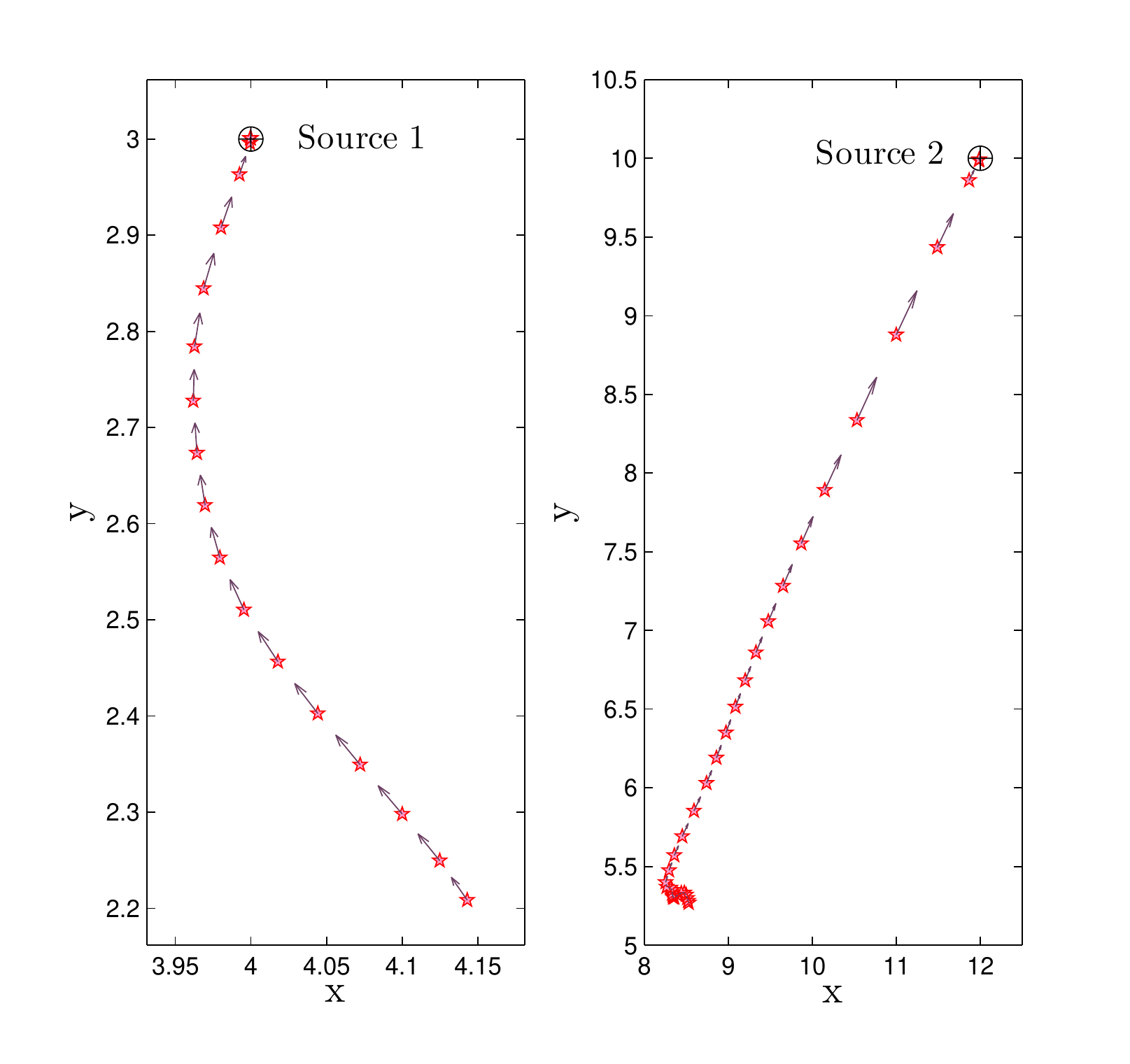}} \subfigure[
]{\includegraphics[width=3.2in ]{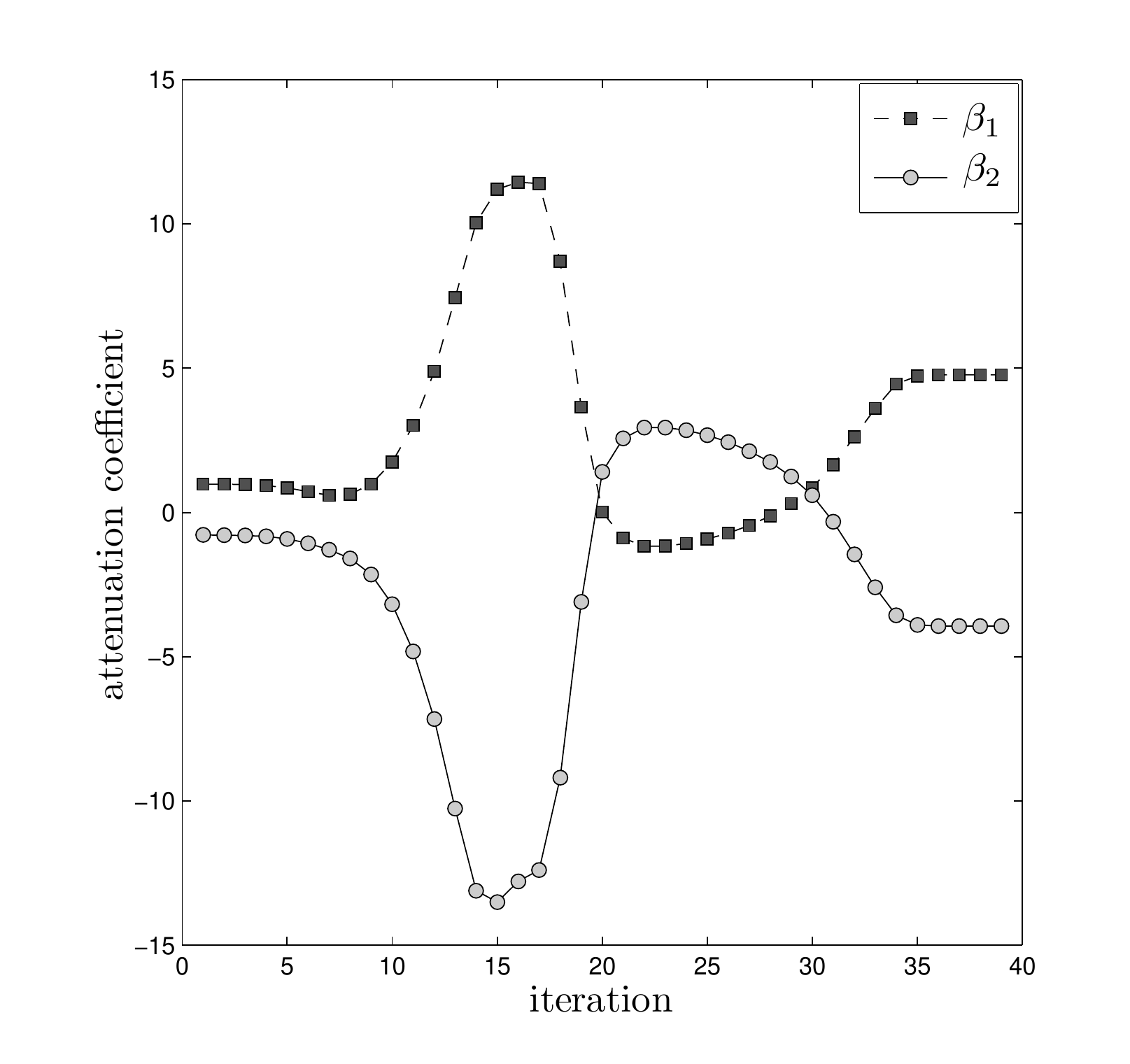} }\hspace{-.5cm}}
\caption{(a) Concurrent estimates of two sources located at
$(4,3)$ and $(12,10)$ (b) The corresponding evolution of the
attenuation model parameters} \label{fig4}
\end{figure*}
As a more challenging problem, we consider concurrent localization
of the two sources located at $(4,3)$ and $(12,10)$. Fig.
\ref{fig4} shows the iterative procedure of finding the sources.
The estimated source locations are $(4.000,3.001)$ and
$(11.989,9.993)$ and the attenuation parameters are estimated to
be $\beta_1=4.77$ and $\beta_2=-3.94$. Again an accurate match
between the exact source locations and the estimated ones is
observable.
\begin{figure*}\centering
\epsfig{file=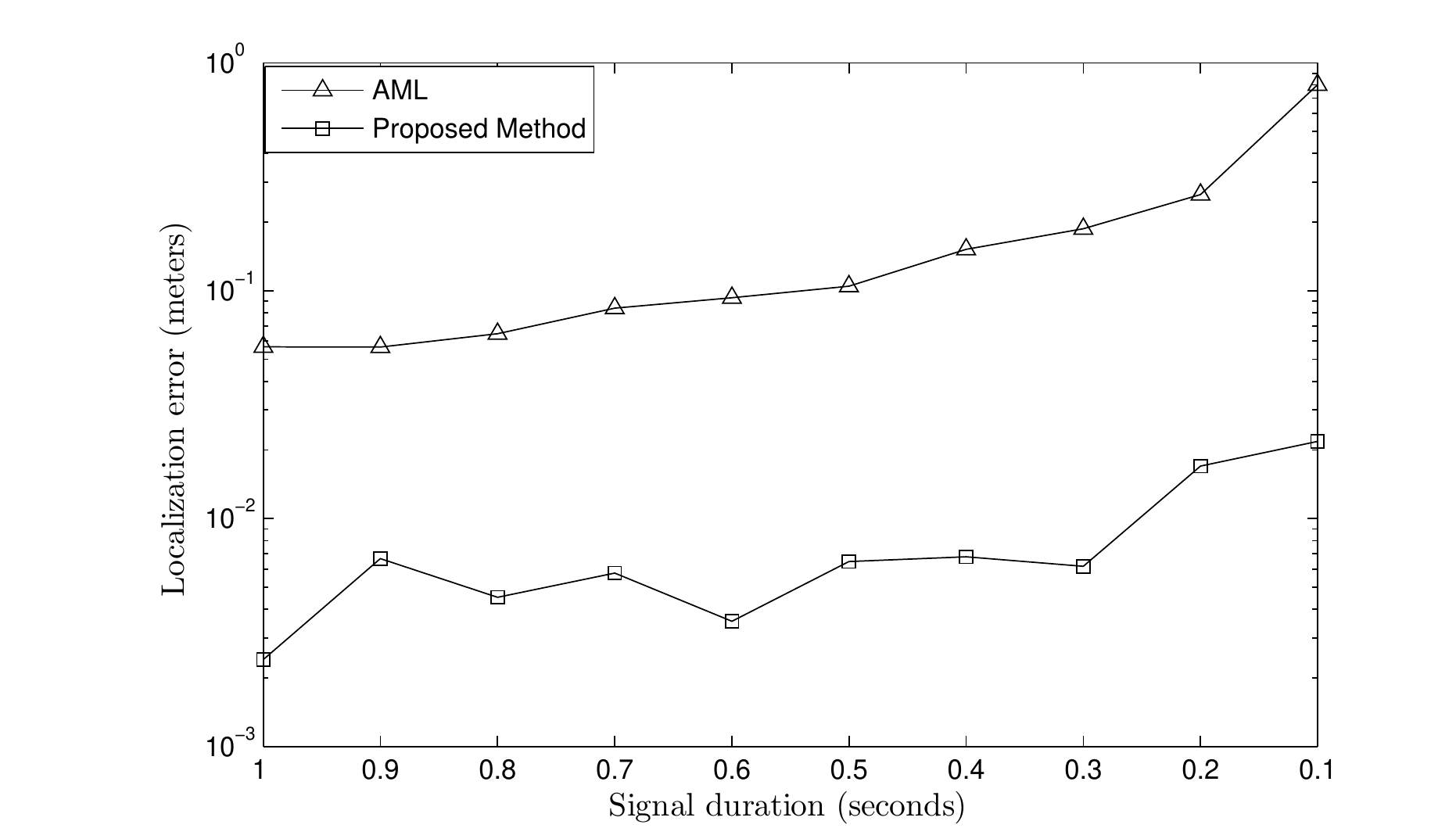,width=.65\linewidth,clip=} \caption{The
localization error verses the signal duration for our proposed
method and the AML method}\label{fig5}
\end{figure*}

We further examine our proposed method through a comparison with
the AML method developed in \cite{chen2002maximum}. For this
purpose we start reducing the signal samples by reducing the
signal duration from 1 to 0.1 seconds and observing the error
caused in the source estimation. Here we consider the single
source localization for the source being located at $(12,10)$.
Fig. \ref{fig5} shows the resulting error as the signal duration
decreases in both methods. As it is clearly observed, using both
time delay and attenuation information helps our method provide
better estimates of the source locations even with less available
data compared to the AML method which only uses the time delay
information. We further examine the performance of both methods
for various SNR values. In Fig. \ref{fig6} the CRLB is calculated
for the same single source scenario with the source located at
$(12,10)$. The RMS errors in estimating the $x$ and $y$ components
of the source are obtained through 50 independent noise
realizations for every SNR value shown in the figure. Again the
proposed method shows an acceptable performance regarding the
closeness to the CRLB and the superior performance compared to the
AML method.

\begin{figure*}
{\subfigure[ ]{\hspace{-.4cm}\includegraphics[width=3.4in ]{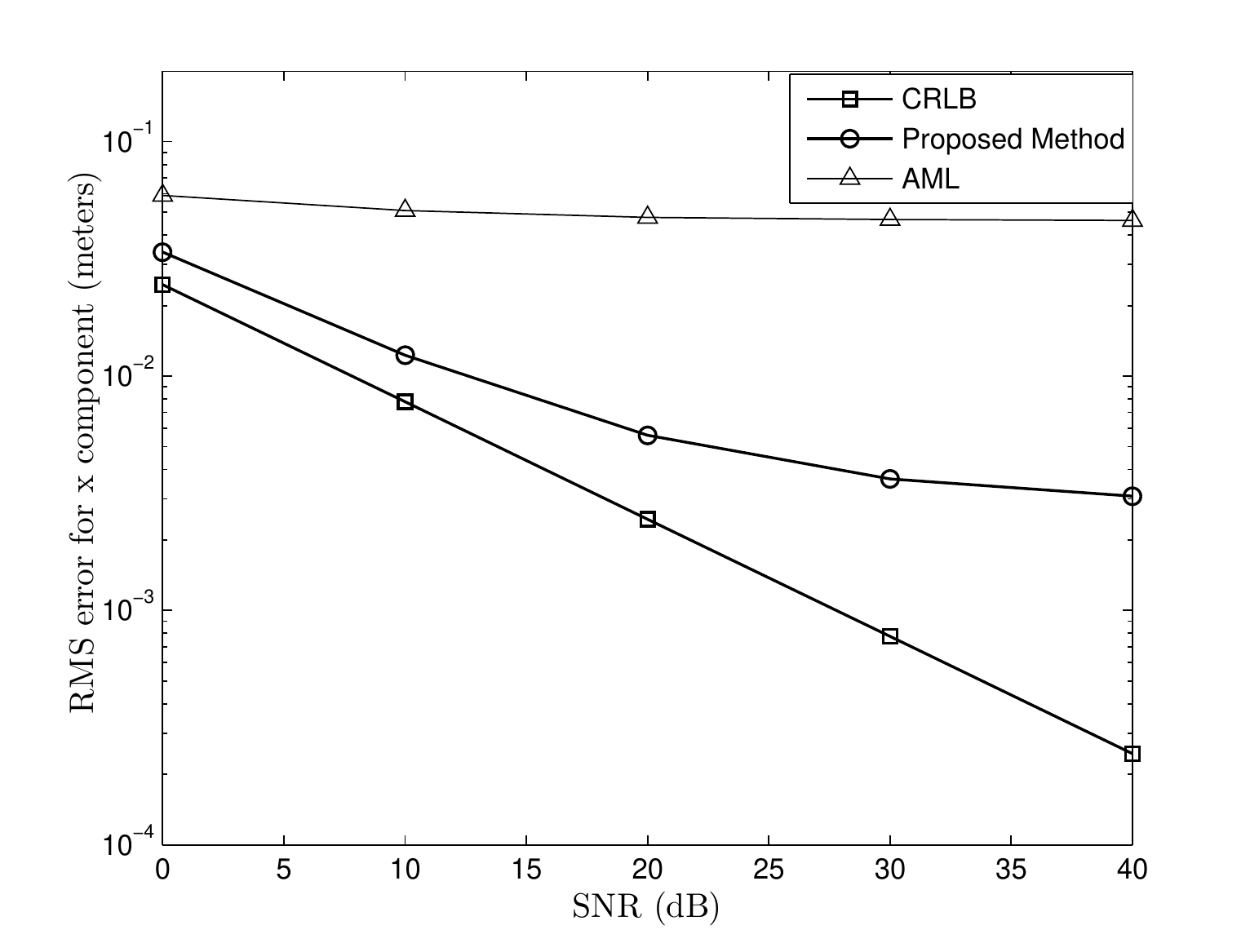}}
\hspace{-.2cm}\subfigure[ ]{\includegraphics[width=3.4in ]{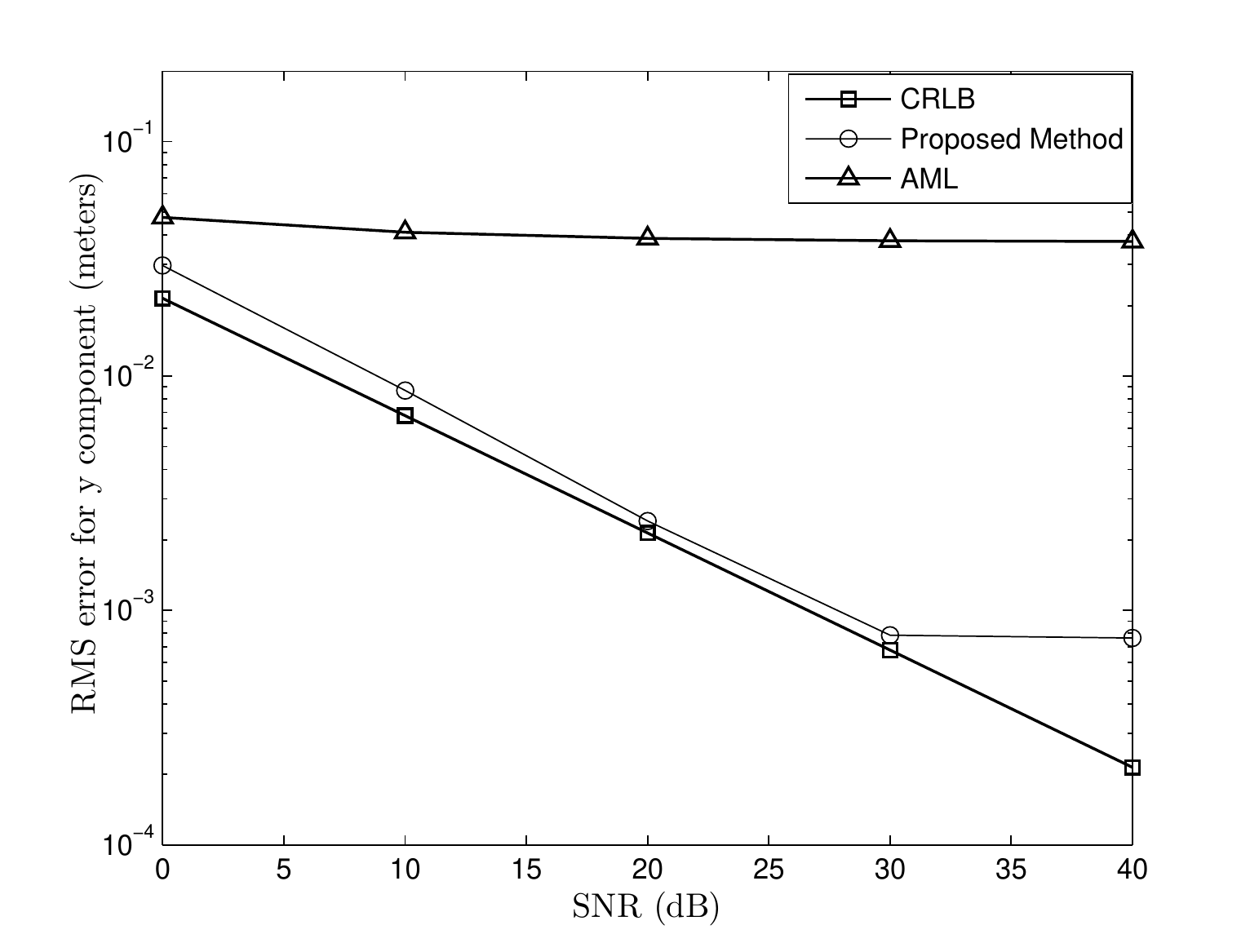}
}} \caption{(a) The RMS error in estimating the $x$ component of a
source located at $(12,10)$. (b) The corresponding RMS error in
estimating the $y$ component of the source.} \label{fig6}
\end{figure*}

\subsection{Example 2}
In a more realistic scenario, we examine the performance of the
proposed method in a noisy environment where sensor
synchronization error and reverberation are likely to happen. The
sensor network configuration is shown in Figure \ref{fig7}, where
three circular arrays each composed of 25 sensors centered at
points (15,5), (2,15) and (5,28) are considered. The acoustic
source is located at (35,25) and the signal specifications are the
same as the previous example. For this example $G_{\max}$ is taken
to be 20 to  benefit more from a global search of a cost function
which may not be as well-behaved as the previous example due to
bringing more unknown parameters into the problem. The low order
attenuation model considered in this example is
$\alpha(\rho)=\rho^{-1}+\beta\rho^{-2}$ with $\beta$ as the tuning
parameter. Again an SNR of 20 dB is considered at all sensors for
all the experiments.
\begin{figure*}\centering
\epsfig{file=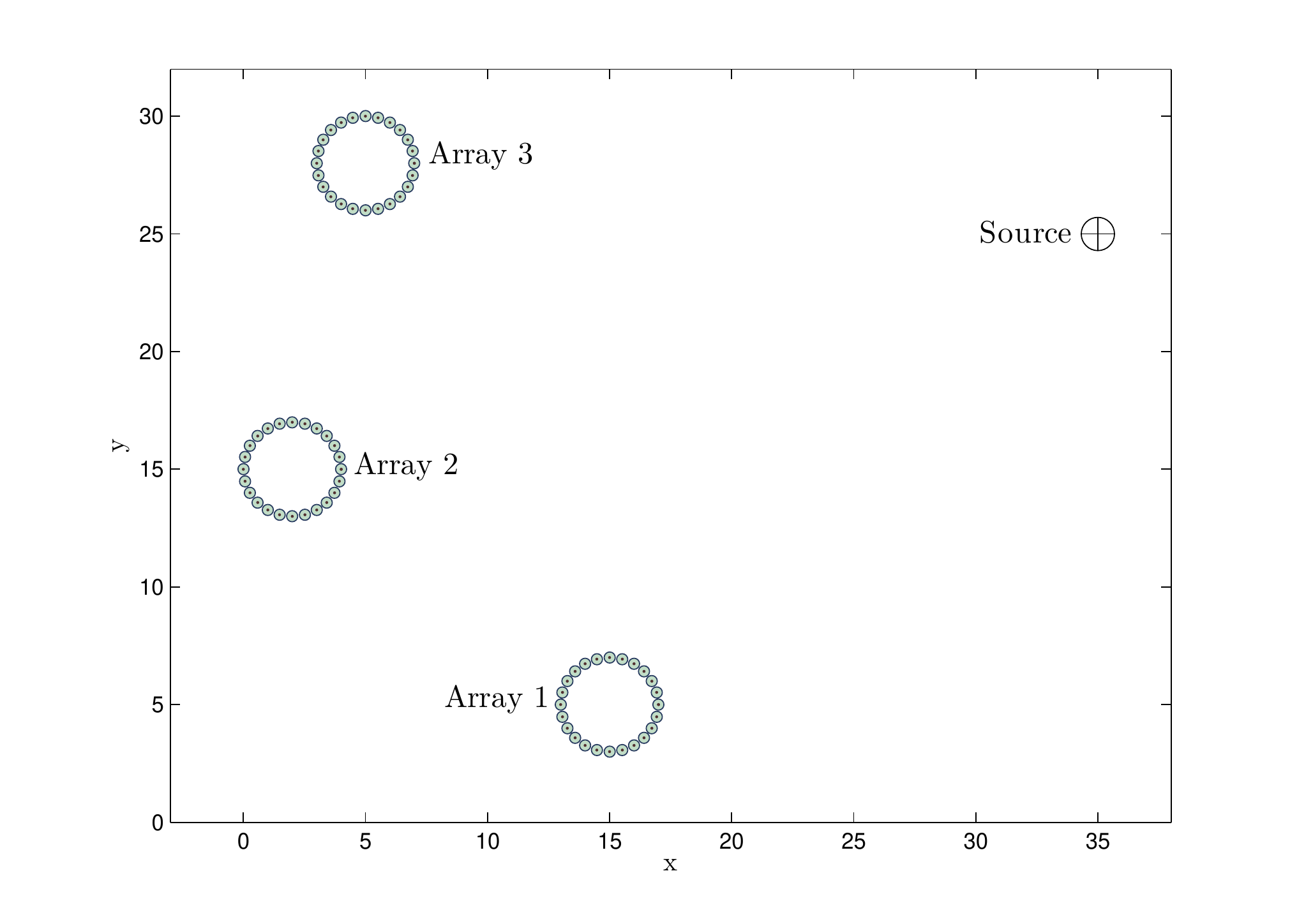,width=.65\linewidth,clip=} \caption{A sensor
network configuration: each array consists of 25 sensors
}\label{fig7}
\end{figure*}

We first examine the case that the sensors are not exactly
synchronized to receive the data. For this purpose we rewrite the
main component of the signal in (\ref{eq1}) as
\begin{align}
\label{eq15}
\sum_{n=1}^{N}\alpha(\rho_{m,n})s_n(t-\tau_{m,n}+\zeta),
\end{align}
where $\zeta$ is a random variable uniformly distributed around
zero. Equation (\ref{eq15}) basically models the asynchronous
measurements of the sensor data. In Table 1 we have provided the
localization results for three different synchronization error
variances $0.5$, $1$ and $2$ milliseconds. Clearly the phase error
is a destructive phenomenon in TDOA localization algorithms,
however, considering the localization errors in Table 1, one would
observe that exploiting the attenuation information beside the
phase information enables our algorithm to perform a rather
accurate localization task in case of sensors being out of
synchronism.

Furthermore, a more challenging problem is when the reverberation
is also taken into account. In theory, for the emitted signal to
arrive at every measuring sensor, an individual multi-path filter
should be considered. Although the formulation in this paper is
general, for the purpose of this example we have made a reasonable
and practical assumption that for all the sensors within each
array, the filter representing the multi-path is identical. In
general the sensor network may be represented as a collection
several clusters each composed of sensors closely placed and each
cluster treated as a single receiving node. This assumption
prevents dealing with a large collection of unknowns
($\gamma_{m,n,p}$ and $\hat\tau_{m,n,p}$) for every source-sensor
pair and aggregates them into fewer parameters each assigned to
the clusters.

To generate a reverberated signal we use the multi-path FIR
filters shown in Figure \ref{fig8} where three or four shifted
scales of the signal are added to it. For the localization
purpose, however, we only consider finding the main indirect path.
In other words, for every array shown in Figure \ref{fig7} only
one multi-path coefficient $\gamma$ and one multi-path delay
$\hat\tau$ is to be estimated which totally brings 6 unknowns
associated with the multi-path phenomenon into the minimization
problem. The remaining minimization unknowns are the source
coordinates and the attenuation coefficients as before. The fourth
row of Table 1 shows the localization result for this problem.
Although the number of unknowns were relatively higher than the
previous examples and the cost function is clearly not as
well-behaved as before, using DE as the initial minimization
scheme provides a suitable starting state for the LMA and this
sequential technique helps the algorithm make a rather accurate
localization in a noisy and reverberated environment. The fifth
row of Table 1 corresponds to the case of having both the
multi-path and the synchronization issues, for which the results
are still promising. The progressive estimates of the target
throughout the minimization are shown in Figure \ref{fig9}.

\begin{table*}[t]\scriptsize
\setlength\arrayrulewidth{1pt}\arrayrulecolor{black}
\caption{Localization Results for the Sensor Network Configuration
in Figure \ref{fig7}.} \label{table2} \centering
    \begin{tabular}{ | c | c | c | c | c |c|c|}\hline
 \multicolumn{2}{|c|}{Type of Problem}& \scriptsize Synchronization& \scriptsize &\scriptsize Number of&\scriptsize Estimated&\scriptsize Localization\\
\cline{1-2}
 \scriptsize Array\normalsize&&\scriptsize Error&$G_{\max}$&\scriptsize LMA&\scriptsize Target&\scriptsize Error\\
     \scriptsize Synchronization \normalsize&\scriptsize
     Reverberation\normalsize&\scriptsize Variance (mS)&&\scriptsize Iterations&\scriptsize Coordinates&\scriptsize (meters)\\
\hline $\blacksquare$&$\square$&0.5&20&21&(34.62 , 24.91) &
0.386\\\hline $\blacksquare$&$\square$&1.0&20&28& (33.99 ,
24.71)&1.053
\\\hline
$\blacksquare$&$\square$&2.0&20&31&(33.79 , 24.66) &1.254
\\\hline
 $\square$&$\blacksquare$&0&20 & 24 & (34.89 , 24.97) & 0.113 \\
 \hline $\blacksquare$ &$\blacksquare$&0.5&20 & 25& (34.60 , 24.89) & 0.418 \\
\hline
    \end{tabular}
\end{table*}\normalsize
\begin{figure*}\centering
\epsfig{file=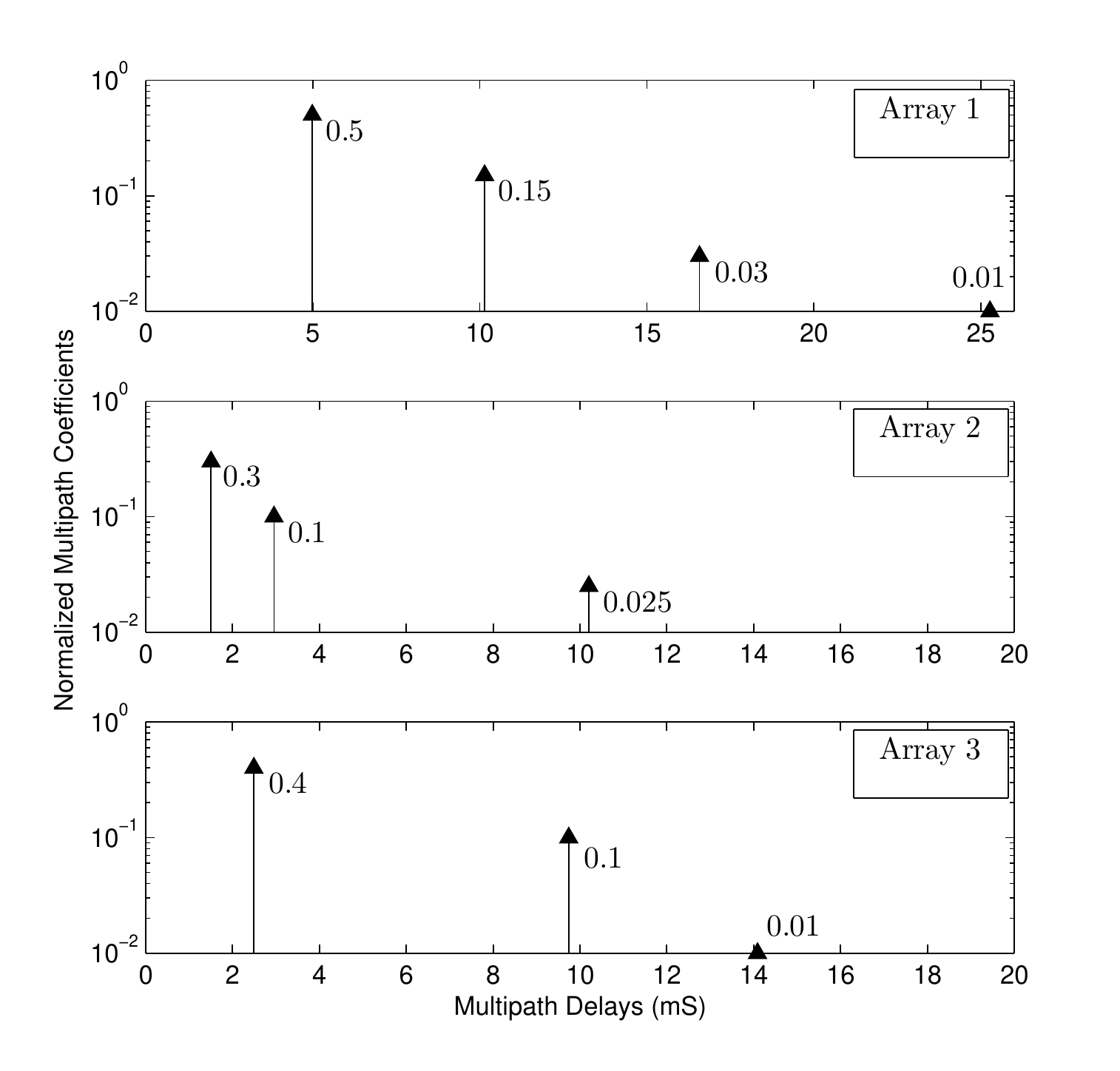,width=.45\linewidth,clip=} \caption{The
multi-path impulse response associated with each array. Every
multi-path model is assumed to hold for all sensors within the
corresponding array. The amplitudes are normalized to the
amplitude of the main signal component.}\label{fig8}
\end{figure*}

\begin{figure*}
{\subfigure[ ]{\hspace{-.4cm}\includegraphics[width=3.4in
]{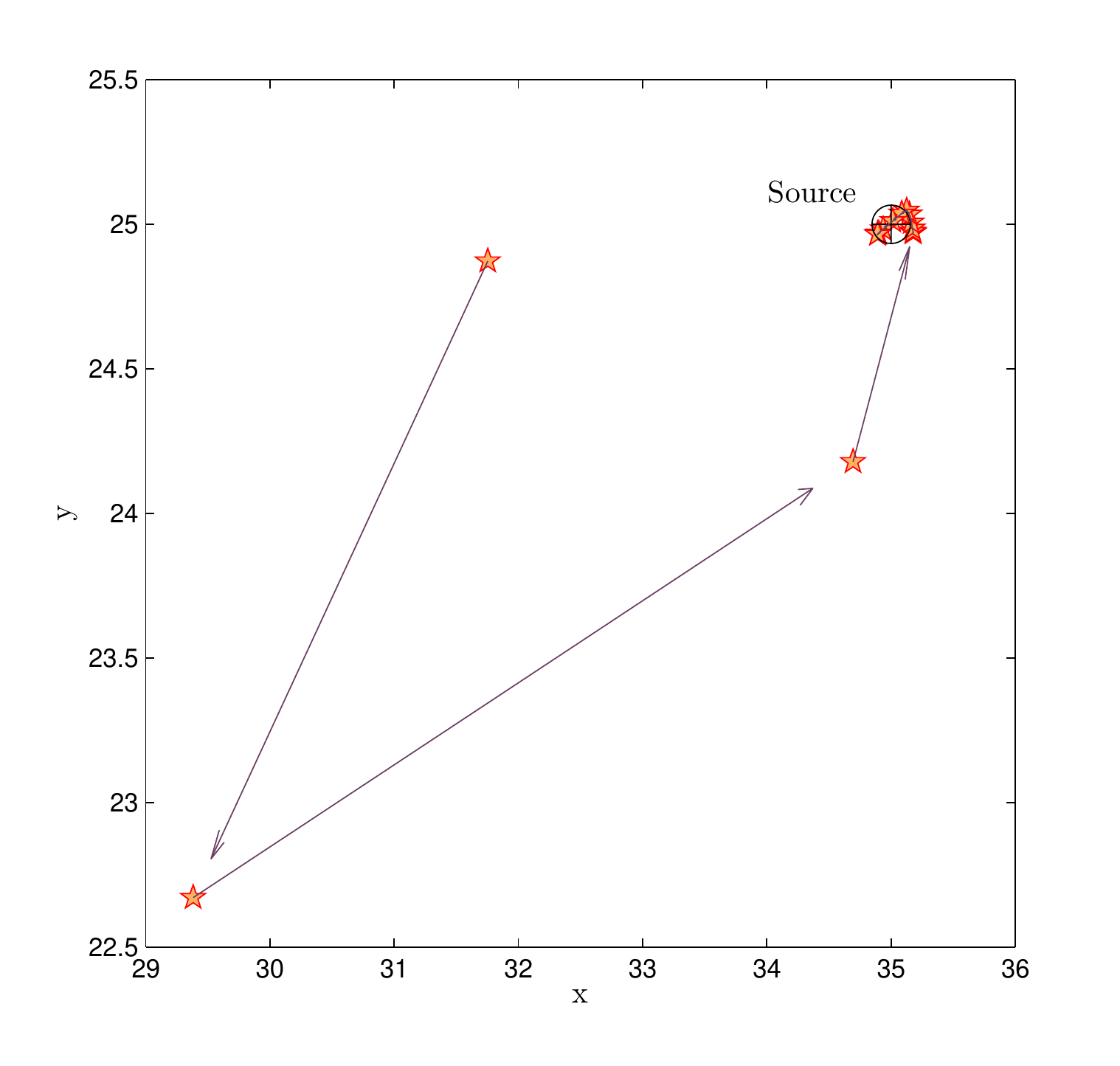}} \hspace{-.2cm}\subfigure[ ]{\includegraphics[width=3.4in
]{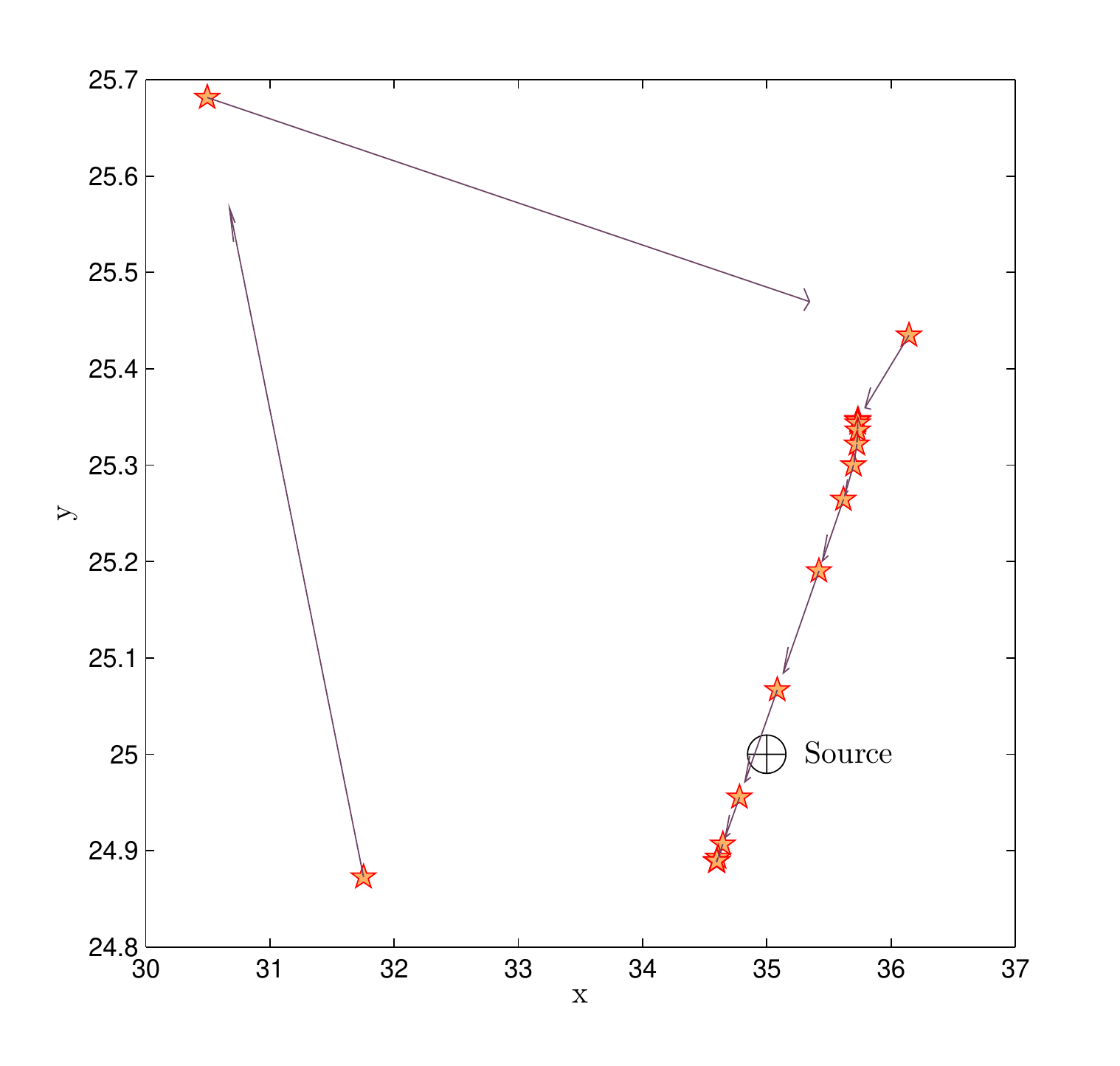} }} \caption{ The progressive estimates of a single source
located at $(35,25)$. The first jumps and good initials correspond
to applying the DE (a) Considering only the multi-path phenomenon
in the environment (b) Considering both, the multi-path and sensor
synchronization error.} \label{fig9}
\end{figure*}

\section{Conclusion} \label{sec6}
In this paper, we proposed an efficient method for localization of
multiple wideband sources based on both signal attenuation and
time delay information. The method developed in this paper models
the localization problem as a minimization problem and provides an
additional flexibility of not being exactly aware of the signal
attenuation model. We propose a certain function space for the
unknown model, and tune it iteratively along to estimate the
signal source locations. The minimization scheme used here is a
hybrid algorithm, combining the differential evolution with the
Levenberg-Marquardt algorithm. This combination increases the
chances of finding a global minima while benefits from the speed
and computational advantages of Newton methods. The accuracy and
performance of the method is examined through several simulations
depicting a noisy environment, a multi-path environment and lack
of synchronization among sensors. In the simulations, we compared
our approach with the approximate maximum likelihood method which
show the superiority of the proposed method.

\section{Appendix}
As mentioned earlier, in order to find columns of the Jacobian, we
are required to find $\partial \boldsymbol{Q}/\partial \theta$,
where $\theta$ is one of the unknown parameters $x_{S_n}$,
$y_{S_n}$, $\beta_\ell$, $\gamma_{m,n,p}$ or $\hat\tau_{m,n,p}$.
Since $\boldsymbol{Q}$ is a vector containing sub-vectors
$\boldsymbol{Q}(f)$ for $f=0,1,\cdots,n_f/2$, we will only find
$\partial \boldsymbol{Q}(f)/\partial \theta$ and clearly forming
$\partial \boldsymbol{Q}/\partial \theta$ would be aligning the
corresponding sub-vectors.

We first start with replacing the pseudo-inverse of
$\boldsymbol{\tilde K}(f)$ in (\ref{eq12}) which states that
\begin{align}
\label{eqa0}\nonumber
\boldsymbol{Q}(f)&=\boldsymbol{X}(f)\\&-\boldsymbol{\tilde K}(f)
\big(\boldsymbol{\tilde K}^H(f)\boldsymbol{\tilde
K}(f)\big)^{-1}\boldsymbol{\tilde K}^H(f)
 \boldsymbol{X}(f).
\end{align}
We can clearly see that finding $\partial
\boldsymbol{Q}(f)/\partial \theta$ requires having $\partial
\boldsymbol{\tilde K}(f) \big(\boldsymbol{\tilde
K}^H(f)\boldsymbol{\tilde K}(f)\big)^{-1}\boldsymbol{\tilde
K}^H(f)/
\partial \theta$. We therefore preliminarily
derive some related equations. Consider a matrix $\boldsymbol{M}$,
not in general rectangular, elements of which are dependent on a
real variable $\theta$. We assume
$\big(\boldsymbol{M}^H\boldsymbol{M}\big)^{-1}$ exists or in other
words
$\boldsymbol{M}^\dag=\big(\boldsymbol{M}^H\boldsymbol{M}\big)^{-1}\boldsymbol{M}^H$.
Using product rule we have
\begin{small}
\begin{align}\label{eqa1}\nonumber
\frac{\partial \boldsymbol{M} \boldsymbol{M}^\dag}{\partial
\theta}&= \frac{\partial }{\partial \theta}\boldsymbol{M}
\big(\boldsymbol{M}^H\boldsymbol{M}\big)^{-1}\boldsymbol{M}^H\\\nonumber
&=\frac{\partial \boldsymbol{M}}{\partial \theta}
\big(\boldsymbol{M}^H\boldsymbol{M}\big)^{-1}\boldsymbol{M}^H +
\boldsymbol{M} \frac{\partial
\big(\boldsymbol{M}^H\boldsymbol{M}\big)^{-1}}{\partial
\theta}\boldsymbol{M}^H\\\nonumber &+ \boldsymbol{M}
\big(\boldsymbol{M}^H\boldsymbol{M}\big)^{-1} \frac{\partial
 \boldsymbol{M}^H}{\partial \theta}\\ &=\frac{\partial \boldsymbol{M}}{\partial
 \theta}\boldsymbol{M}^{\dag} \!+\! \boldsymbol{M}
\frac{\partial
\big(\boldsymbol{M}^H\boldsymbol{M}\big)^{-1}}{\partial
\theta}\boldsymbol{M}^H\!+\!{\boldsymbol{M}^{\dag}}^H\frac{\partial
 \boldsymbol{M}^H}{\partial \theta}
\end{align}
\end{small}
Also we know that for an invertible matrix ${\boldsymbol{\tilde
M}}$ again dependent on $\theta$ we have
\begin{equation}
\label{eqa2} \frac{\partial{\boldsymbol{\tilde M}}^{-1}}{\partial
\theta} =-{\boldsymbol{\tilde M}}^{-1}\frac{\partial
{\boldsymbol{\tilde M}}}{\partial \theta}{\boldsymbol{\tilde
M}}^{-1}.
\end{equation}
Using (\ref{eqa2}) in (\ref{eqa1}) regarding the term $\partial
\big(\boldsymbol{M}^H\boldsymbol{M}\big)^{-1}/\partial \theta$,
would result in
\begin{align}\label{eqa3}\nonumber
\frac{\partial \boldsymbol{M} \boldsymbol{M}^\dag}{\partial
\theta}&=
(\boldsymbol{I}-{\boldsymbol{M}^{\dag}}^H\boldsymbol{M}^H)\frac{\partial
\boldsymbol{M}}{\partial
\theta}\boldsymbol{M}^{\dag}\\&+{\boldsymbol{M}^{\dag}}^H\frac{\partial
\boldsymbol{M}^H}{\partial
\theta}(\boldsymbol{I}-\boldsymbol{M}\boldsymbol{M}^{\dag}).
\end{align}
Based on (\ref{eqa0}), and knowing (\ref{eqa3}), we now have
\begin{equation}
\label{eqa4} \frac{\partial \boldsymbol{Q}(f)}{\partial \theta} =
\big( \boldsymbol{P}(f) +
\boldsymbol{P}^H(f)\big)\boldsymbol{X}(f),
\end{equation}
where
\begin{equation}
\label{eqa5} \boldsymbol{P}(f)=\Big({\boldsymbol{\tilde
K}^\dag}^H(f)\boldsymbol{\tilde
K}^H(f)-\boldsymbol{I}\Big)\frac{\partial \boldsymbol{\tilde
K}(f)}{\partial \theta}\boldsymbol{\tilde K}^\dag(f).
\end{equation}

To complete the derivation we only need to have $\partial
\boldsymbol{\tilde K}(f)/\partial \theta$, which is already
discussed in Section \ref{secCRLB}.
{\ifthenelse{\boolean{publ}}{\footnotesize}{\small}
 \bibliographystyle{bmc_article}  
  \bibliography{references} }     


\ifthenelse{\boolean{publ}}{\end{multicols}}{}
\end{bmcformat}
\end{document}